\documentclass[a4paper,conference]{IEEEtran}

\usepackage[left=1.57cm,right=1.57cm,top=0.95cm,bottom=2.54cm]{geometry}

\RequirePackage{ifthen}
\newboolean{debugLinks}\setboolean{debugLinks}{false}
\newboolean{debugCites}\setboolean{debugCites}{false}
\newboolean{debugPages}\setboolean{debugPages}{false}

\pdfpageattr{/Group << /S /Transparency /I true /CS /DeviceRGB >>}

\usepackage[utf8]{inputenc}
\usepackage[T1]{fontenc}
\usepackage{textcomp}
\usepackage{microtype}
\usepackage{calc}
\usepackage{acro}
\usepackage{comment}

\usepackage[normalem]{ulem}
\usepackage{footmisc}

\usepackage{lipsum}

\usepackage[range-phrase=--,per-mode=symbol-or-fraction,range-units=single,list-units=single,detect-all,forbid-literal-units=true]{siunitx}
\AtBeginDocument{
\DeclareSIUnit\bit{bit}
\DeclareSIUnit\byte{Byte}
\DeclareSIUnit\decibeli{dBi}
\DeclareSIUnit\decibelm{dBm}
\DeclareSIUnit\kmh{\kilo\meter\per\hour}
\DeclareSIUnit\mbps{\mega\bit\per\second}
\DeclareSIUnit\mph{mph}
\DeclareSIUnit\mw{\milli\watt}
\DeclareSIUnit\resourceblock{RB}
\DeclareSIUnit\vehicle{veh}
\DeclareSIUnit\watthour{Wh}
\DeclareSIUnit\usdollar{US\$}
\DeclareSIUnit\pp{\ac{PP}}
}

\usepackage{silence}
\usepackage{csquotes}
\ifthenelse{\boolean{debugCites}}{
    \usepackage[backend=biber,style=verbose,bibstyle=ieee,doi=true,isbn=false,mincitenames=1,maxcitenames=2,minbibnames=1,maxbibnames=6]{biblatex}
}{
    \usepackage[backend=biber,style=ieee,bibstyle=ieee,doi=false,isbn=false,mincitenames=1,maxcitenames=2,minbibnames=1,maxbibnames=6]{biblatex}
}
\DeclareFieldFormat{sentencecase}{#1} 
\DeclareFieldFormat{titlecase}{#1} 
\usepackage{xpatch}
\xpatchbibmacro{textcite}{\addspace}{\addnbspace}{}{}
\xpatchbibmacro{Textcite}{\addspace}{\addnbspace}{}{}
\DefineBibliographyStrings{english}{
    andothers = et~al\adddot\addspace
}

\DeclareSourcemap{
    \maps[datatype=bibtex]{
        \map[overwrite=true]{
            \step[fieldsource=school,fieldtarget=institution]
        }
    }
}
\RenewDocumentCommand\cite{m}{%
    \autocite{#1}
}

\usepackage{amsmath}

\usepackage{amssymb}
\usepackage{amsfonts}
\usepackage{amsthm}

\usepackage{booktabs}
\usepackage[]{url}
\makeatletter
\def\url@cmsstyle{%
  \def\UrlSpecials{%
    \do\-{\mathchar`-}%
  }%
}
\makeatother

\usepackage[inline]{enumitem}

\usepackage[caption=false,font=footnotesize]{subfig}
\usepackage[pdftex]{graphicx}
\DeclareGraphicsExtensions{.pdf,.png,.jpg,.jpeg}
\usepackage{xfrac}

\usepackage{tikz}
\usetikzlibrary{arrows}
\usetikzlibrary{calc}
\usetikzlibrary{chains}
\usetikzlibrary{scopes}

\ifthenelse{\boolean{debugCites} \OR \boolean{debugLinks} \OR \boolean{debugPages}}{
    \WarningsOff[everypage]
    \usepackage[placement=top]{background}
    \SetBgContents{draft}
    \SetBgScale{1}
    \SetBgVshift{-7}
}{
}

\usepackage[american]{babel}
\usepackage{hyphenat}

\usepackage{algorithmic}
\usepackage{algorithm}

\ifthenelse{\boolean{debugLinks}}{
    \usepackage[breaklinks=true]{hyperref}
}{
}

\usepackage[capitalize,noabbrev,nameinlink]{cleveref}
\crefformat{footnote}{#2\textsuperscript{\normalfont #1}#3}
\crefname{Assumption}{Assumption}{Assumptions}
\usepackage{footmisc}

\RequirePackage{xstring}
\RequirePackage{xparse}
\RequirePackage[]{acro}

\makeatletter
\@ifpackagelater{acro}{2019/02/17}{
    \@ifpackagelater{acro}{2020/04/29}{
        \NewDocumentCommand\acrodef{mO{#1}mG{}}{\DeclareAcronym{#1}{short={#2}, long={#3}, #4}}
    }{
        \NewDocumentCommand\acrodef{mO{#1}mG{}}{\DeclareAcronym{#1}{short={#2}, long={#3}, foreign-plural={}, #4}}
    }
}{
    \NewDocumentCommand\acrodef{mO{#1}mG{}}{\DeclareAcronym{#1}{short={#2}, long={#3}, #4}}
}
\makeatother

\makeatletter
\@ifpackagelater{acro}{2020/04/29}{
    \NewAcroTemplate{inparen}{%
        \acroiffirstTF{%
            \acrowrite{long}%
            \acspace%
            \scalebox{0.5}[1]{(}%
            \acroifT{foreign}{\acrowrite{foreign}, }%
            \acrowrite{short}%
            \acroifT{alt}{ \acrotranslate{or} \acrowrite{alt}}%
            \acrogroupcite
            \scalebox{0.5}[1]{)}%
        }{%
            \acrowrite{short}%
        }%
    }
    \NewAcroTemplate{paren}{%
        \acroiffirstTF{%
            \acrowrite{long}%
            ,%
            \acspace%
            \acroifT{foreign}{\acrowrite{foreign}, }%
            \acrowrite{short}%
            \acroifT{alt}{ \acrotranslate{or} \acrowrite{alt}}%
            \acrogroupcite
        }{%
            \acrowrite{short}%
        }%
    }
}{
    \DeclareAcroFirstStyle{inparen}{inline}
      {
        brackets = true,
        brackets-type = {%
          \scalebox{0.5}[1]{(}%
        }{%
          \scalebox{0.5}[1]{)}%
        }
      }
}
\makeatother

\acrodef{AoI}{Age of Information}
\acrodef{ANRM}{All-or-Nothing Receiver Model}
\acrodef{AV}{Autonomous Vehicle}
\acrodef{BW}{Bandwidth}
\acrodef{BSM}{Basic Safety Message}
\acrodef{BLER}{Block-Error Rate}
\acrodef{C-ITS}{Cooperative Intelligent Transport Systems}
\acrodef{CAM}{Cooperative Awareness Message}
\acrodef{CBR}{Channel Busy Ratio}
\acrodef{CCDF}{Complementary Cumulative Distribution Function}
\acrodef{CLR}{Collision Loss Ratio}
\acrodef{CST}{Collision State Time}
\acrodef{CR}{Channel occupancy Ratio}
\acrodef{COV}{Coefficient of Variation}
\acrodef{C-V2X}{Cellular V2X}
\acrodef{DCC}{Decentralized Congestion Control}
\acrodef{DDPM}{Distance-Dependent Propagation Model}
\acrodef{DRL}{Deep Reinforcement Learning}
\acrodef{DS}{Dynamic Scheduling}
\acrodef{DST}{Disconnection State Time}
\acrodef{DSRC}{Dedicated Short-Range Communication}
\acrodef{DTMC}{Discrete Time Markov Chain}
\acrodef{FD}{Full Duplex}
\acrodef{GP}{Geometric persistence}
\acrodef{HARQ}{Hybrid Automatic Repeat Request}
\acrodef{i.i.d.}{independent and identically distributed}
\acrodef{IoT}{Internet of Things}
\acrodef{ITS}{Intelligent Transportation Systems}
\acrodef{LOS}{Line of Sight}
\acrodef{MAC}{Medium Access Control}
\acrodef{MCS}{Modulation and Coding Scheme}
\acrodef{NR}{New Radio}
\acrodef{NOMA}{Non-Orthogonal Multiple Access}
\acrodef{PAoI}{Peak Age of Information}
\acrodef{P}{probability of persistence}
\acrodef{PER}{Packet Error Rate}
\acrodef{PCR}{Packet Collision Ratio}
\acrodef{PDF}{Probability Distribution Function}
\acrodef{PDB}{Packet Delay Budget}
\acrodef{PDR}{Packet Delivery Ratio}
\acrodef{PIR}{Packet Inter-Reception Delay}
\acrodef{PLR}{Propagation Loss Ratio}
\acrodef{PPP}{Poisson Point Process}
\acrodef{PRR}{Packet Reception Ratio}
\acrodef{PSSCH}{Physical sidelink Shared Channel}
\acrodef{PSFCH}{Physical sidelink Feedback Channel}
\acrodef{RB}{Resource Block}
\acrodef{RC}{Reselection Counter}
\acrodef{RRI}{Resource Reservation Interval}
\acrodef{rxSensitivity}{Receiver sensitivity}
\acrodef{RSSI}{Received Signal Strength Indicator}
\acrodef{RSRP}{Reference Signal Receive Power}
\acrodef{OFDM}{Orthogonal Frequency-Division Multiple Access}
\acrodef{OMA}{Orthogonal Multiple Access}
\acrodef{SC}{Sub-Channel}
\acrodef{SIC}{Successive Interference Cancellation}
\acrodef{SL}{sidelink}
\acrodef{RCS}{Repeated Contention Scheduling}
\acrodef{SDR}{Software Defined Radio}
\acrodef{SCS}{Sub-Carrier Spacing}
\acrodef{SCI}{sidelink Control Information}
\acrodef{SNR}{Signal-to-Noise Ratio}
\acrodef{SNIR}{Signal-to-Noise-plus-Interference Ratio}
\acrodef{SPS}{Semi-Persistent Scheduling}
\acrodef{TB}{Transport Block}
\acrodef{TDB}{Time Delay Budget}
\acrodef{UE}{User Equipment}
\acrodef{VoI}{Value of Information}
\acrodef{V2X}{Vehicle-to-Everything}
\acrodef{V2V}{Vehicle-to-Vehicle}
\acrodef{WAVE}{Wireless Access in Vehicular Environments}
\acrodef{WBS}{Wireless Blind Spot}

\newcommand{\useacronyms}{
}
\xapptocmd{\maketitle}{\useacronyms}{}{}
\xapptocmd{\acresetall}{\useacronyms}{}{}
\xapptocmd{\printbibliography}{\acuseall}{}{} 

\newcommand{\sectionacronyms}{
}
\AddToHook{cmd/section/before}{\sectionacronyms{}}

\usepackage{todonotes}

\usepackage[noadjust,noparboxrestore]{marginnote}
\makeatletter
\def\todo{%
    \begingroup%
    \color{magenta}%
    \ifnum\@floatpenalty<0\relax%
    \else%
        \setlength{\columnsep}{2cm} 
        \marginnote{\color{magenta}\rule{2pt}{1em}}%
        \obeylines%
        \begingroup\lccode`~=`\^^M\lowe\ac{RCS}se{\endgroup\def~}{\par\leavevmode}%
        \parindent0em%
        \catcode`\_=\active%
        \catcode`\<=\active\lccode`~=`<\lowe\ac{RCS}se{\def~}{$<$}%
        \catcode`\>=\active\lccode`~=`>\lowe\ac{RCS}se{\def~}{$>$}%
        \catcode`\#=\active\lccode`~=`\#\lowe\ac{RCS}se{\def~}{$\#$}%
        \catcode`\^=\active\lccode`~=`\^\lowe\ac{RCS}se{\def~}{$\hat{~}$}%
        \catcode`\&=\active\lccode`~=`\&\lowe\ac{RCS}se{\def~}{\&}%
    \fi%
    \todoCtd%
}\def\todoCtd#1{%
    TODO: #1%
    \ifx&#1&..\fi%
    \ifnum\@floatpenalty<0\relax%
    \else%
    \fi%
    \endgroup%
    \relax%
}
\makeatother

\NewDocumentCommand\IEEE{ s m >{\SplitArgument{4}{/}}d[] }{%
    \IfBooleanTF{#1}{}{IEEE\,}
    \nolinebreak[2]
    #2%
    \IfNoValueTF{#3}{%
    }{%
        \sommerIEEELettersSlashed#3%
    }%
}
\newcommand{\sommerIEEELettersSlashed}[5]{%
    \IfNoValueTF{#2}{%
    }{%
        \nolinebreak[3]
    }%
    #1%
    \IfNoValueTF{#2}{}{/#2}%
    \IfNoValueTF{#3}{}{/#3}%
    \IfNoValueTF{#4}{}{/#4}%
    \IfNoValueTF{#5}{}{/#5}%
}

\usepackage{listings}

\begin{document}
\begin{figure*}[]
\centering
\noindent\fbox{%
\begin{minipage}{\dimexpr\linewidth-2\fboxsep-2\fboxrule\relax}
\footnotesize
\noindent \copyright\ 2026 IEEE. Personal use of this material is permitted. Permission from IEEE must be obtained for all other uses, in any current or future media, including reprinting/republishing this material for advertising or promotional purposes, creating new collective works, for resale or redistribution to servers or lists, or reuse of any copyrighted component of this work in other works.

\vspace{0.15cm}
\hrule height 0.4pt 
\vspace{0.15cm}

\noindent \textit{2026 Mediterranean Artificial Intelligence and Networking Conference (MAIN 2026)}
\end{minipage}}
\end{figure*}

\title{Repeated Contention Scheduling: A Novel Resource Allocation Algorithm Toward 6G Vehicular Networks}

\author{%
\IEEEauthorblockN{%
    Alexey Rolich\IEEEauthorrefmark{1},
    Marco Tricco\IEEEauthorrefmark{1},
    Simone Paroli\IEEEauthorrefmark{1},
    Mert Yildiz\IEEEauthorrefmark{1}, and
    Andrea Baiocchi\IEEEauthorrefmark{1}\\%
}%
\IEEEauthorblockA{%
    \IEEEauthorrefmark{1}University of Rome Sapienza, Italy\\%
    \texttt{\{alexey.rolich, mert.yildiz, andrea.baiocchi\}@uniroma1.it}\\%
    \texttt{\{paroli.1853547, tricco.1894220\}@studenti.uniroma1.it}%
}%
}

\markboth{Journal of \LaTeX\ Class Files,~Vol.~14, No.~8, August~2021}%
{Shell \MakeLowe\ac{RCS}se{\textit{et al.}}: A Sample Article Using IEEEtran.cls for IEEE Journals}

\IEEEpubid{0000--0000/00\$00.00~\copyright~2021 IEEE}

\maketitle

\begin{abstract}
Efficient decentralized resource allocation remains a fundamental challenge in NR-V2X sidelink communications, where conventional Semi-Persistent Scheduling (SPS) and Dynamic Scheduling (DS) suffer from persistent collisions and limited adaptability under dynamic and dense conditions. This paper proposes Repeated Contention Scheduling (RCS), a novel resource allocation algorithm based on multi-round, feedback-driven contention that eliminates long-term reservations and enables fully distributed operation. Simulations demonstrate that RCS outperforms SPS and DS in terms of success probability, collision and loss reduction, and timeliness metrics such as Packet Inter-Reception Delay and Age of Information, particularly under high load. The practical feasibility of the approach is validated through an SDR-based experimental testbed, which confirms robust operation under realistic hardware impairments and closely matches theoretical and simulation results. These findings establish RCS as a viable and scalable solution for resource allocation in 5G NR sidelink and a promising candidate for future 6G vehicular communication systems.
\end{abstract}


%

\section{Introduction}
\label{sec:intro}

\ac{ITS} constitute a key component of next-generation transportation systems, enabling improved safety, traffic efficiency, and automation through pervasive connectivity.
Vehicular networks provide the communication backbone of \ac{ITS}, supporting real-time information exchange among vehicles, infrastructure, and vulnerable road users. 
In this context, 5G \ac{NR}-\ac{V2X} has emerged as a technological solution offering high-reliability, low-latency communication tailored to advanced vehicular applications~\cite{garcia2021tutorial}. 
A central feature of \ac{NR}-\ac{V2X} is sidelink communication, which enables direct and decentralized interactions in both in-coverage (Mode~1) and out-of-coverage (Mode~2) scenarios~\cite{garcia2021tutorial,ali2021mode2}.

Efficient sidelink resource allocation remains a fundamental challenge in dynamic vehicular environments characterized by high mobility, heterogeneous traffic density, and rapidly varying channel conditions. 
The 5G \ac{NR}-\ac{V2X} sidelink framework defines two operational modes. 
In Mode~1, the gNB performs centralized scheduling through dynamic grants for per-transmission allocation and configured grants for semi-persistent reservations, supporting both periodic and aperiodic traffic under network coverage. 
In Mode~2, user equipments autonomously select resources based on sensing mechanisms. 
Two standardized allocation schemes are defined for this mode: \ac{SPS} and \ac{DS}~\cite{3GPP_TS_38.321,3GPP_TS_37.985}.

\ac{SPS} adopts a reservation-based strategy in which resources are selected and retained over multiple transmission intervals according to a predefined \ac{RRI}, controlled by parameters such as the \ac{RC} and persistence probability $P_{pers}$. 
In contrast, \ac{DS} follows a non-persistent approach, where resources are selected independently for each packet, making it more suitable for aperiodic traffic~\cite{lusvarghi2023comparative}. 
Although both schemes are standardized, no explicit guideline is provided for their selection, and their effectiveness depends on traffic characteristics and network conditions. 
%

\ac{SPS} is widely adopted due to its efficiency for periodic traffic and reduced control overhead in decentralized operation. 
However, its persistence-based design introduces inherent limitations. 
Repeated reuse of reserved resources increases the probability of persistent packet collisions when multiple vehicles select overlapping resources, particularly in the presence of hidden terminals and sensing inaccuracies~\cite{bazzi2020blindspots}. 
These collisions may persist over consecutive transmissions~\cite{bazzi2020blindspots,ROLICH2024,rolich2023}, leading to inefficient spectrum utilization~\cite{lusvarghi2023comparative} and degraded quality of service under high mobility or network congestion~\cite{rolich2025rethink}.

To mitigate these issues, numerous enhancements have been proposed. 
At the protocol level, 3GPP introduces mechanisms such as short-term sensing~\cite{Ha2022}, resource re-evaluation~\cite{todisco_2024,molinagalan_2024,molinagala_2026}, pre-emption~\cite{Han2024}, and inter-UE coordination~\cite{Shehata2021}, along with feedback-based reliability features including \ac{HARQ} and \ac{PSFCH}~\cite{an2023,mahabal2024,fu2025}. 
From an algorithmic perspective, optimization-based approaches, including convex optimization, game theory, and graph-based formulations, have been applied to improve resource selection~\cite{parvini2024}. 
In parallel, machine learning approaches, particularly reinforcement learning~\cite{ali2026,Hedge2023,li2025,montano2026} and multi-agent systems~\cite{saad2025,xiaolu2025}, enable adaptive and context-aware decision-making~\cite{rolich2026dt,rolich2025tradeoff,shen2025,peng2021,cao2022,Rolich_2024_DCC,Rolich_26_VTM} in decentralized environments. 
Additional directions include the integration of \ac{NOMA} for enhanced spectral efficiency and context-aware strategies such as RSU-assisted or position-based allocation~\cite{bazzi2023,todisco2023_NOMA,hirai2022,fuchikami2024}.

Despite these advances, most approaches remain evolutionary and retain the semi-persistent allocation paradigm, thereby remaining susceptible to recurring collisions. 
Consequently, robust distributed resource allocation remains an open problem. 
Survey and tutorial studies on resource allocation in 5G NR-V2X and beyond~\cite{Shin2023,AlNajjar_2024,Annu2024,Wael2025} confirm that recent progress has largely focused on incremental refinements of \ac{SPS}/\ac{DS}-based mechanisms rather than fundamentally new designs. 
As a result, persistent collisions caused by hidden terminals, sensing inaccuracies, and limited coordination remain unresolved. 
Moreover, non-persistent allocation strategies have received comparatively limited attention, particularly those incorporating effective feedback and coordination without relying on long-term reservations. 
This highlights that it is worth investigating alternative paradigms that overcome the inherent limitations of reservation-based schemes.

Motivated by these limitations, this paper proposes a new sidelink resource allocation approach, termed \ac{RCS}, for \ac{NR}-\ac{V2X} and future 6G-V2X systems. 
Unlike \ac{SPS}, the proposed method \textit{is not based on semi-persistent reservations}. 
Instead, it employs repeated contention rounds~\cite{Zame_2013,Baiocchi_2017,baiocchi2019good}, where vehicles contend for resources for each message they have to send. 
Contention exploits sub-carrier signaling with simultaneous transmission and reception.
The feasibility of this key function is proved via an experimental test-bed.
This design eliminates long-term reservations and replaces them with a dynamically re-evaluated contention process. 
\ac{RCS} is fully compatible with 5G \ac{NR}-\ac{V2X} sidelink while improving reliability and timeliness in dense scenarios, as highlighted by numerical results. 
\ac{RCS} provides a scalable solution for current vehicular networks and a foundation for future intelligent resource allocation mechanisms in 6G vehicular systems.

Main contribution of the paper:
\begin{itemize}
\item We propose a novel resource allocation algorithm, termed \ac{RCS}, based on a repeated contention mechanism for \ac{NR}-\ac{V2X} sidelink communication. Unlike conventional \ac{DS} and \ac{SPS} schemes, the proposed approach eliminates semi-persistent reservations and instead relies on feedback-driven contention rounds, enabling dynamic and collision-resilient resource selection.
\item We demonstrate through extensive simulations that \ac{RCS} outperforms traditional \ac{DS} and \ac{SPS} in terms of reliability, collision mitigation, information timeliness, particularly in dense vehicular scenarios.
\item We validate the practical applicability of \ac{RCS} through a real-world experimental setup, providing a proof-of-concept implementation that confirms its feasibility and effectiveness under realistic deployment conditions.
\end{itemize}

The rest of this paper is organized as follows.
\Cref{sec:algorithm} presents the proposed \ac{RCS} algorithm, detailing its repeated contention mechanism and design principles.
\Cref{sec:sim} describes the simulation framework, including system assumptions, parameter configuration, and evaluation metrics relevant to reliability and timeliness. Also, this section reports the simulation results and provides a comparative performance analysis against baseline schemes.
\Cref{sec:experiment} presents the experimental proof-of-concept implementation and discusses practical considerations and observed performance in a real-world setup, as illustrated in the prototype system.
Finally, \Cref{sec:conclusion} summarizes the main findings and outlines directions for future research.

\section{Description of \ac{RCS}}
\label{sec:algorithm}

In this section, we describe the system model and the configuration of 5G NR multiple access to support \ac{RCS}.
Then, the \ac{RCS} algorithm is presented in \cref{subsec:ReCoalgo}.
Then we provide an analysis of \ac{RCS} (\cref{subsec:ReCoanalysis}).

\subsection{System model}
\label{subsec:sysdescr}

Let us consider $n$ nodes, each one generating update messages.
We assume a message fits into a single \ac{SC}.
The average generation time, i.e., the mean time elapsing for a node to generate a new update message, is denoted with $T_g$.

\begin{figure}[t]
\centering\includegraphics[width=0.42\textwidth]{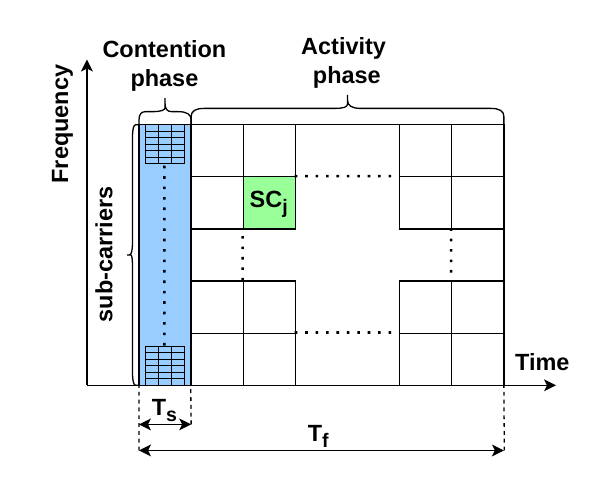}
\vspace{-0.5cm}
\caption{Configuration of the channel using \ac{RCS}.}
\label{fig:RCS_example}
\vspace{-0.2cm}
\end{figure}

The multiple access structure is shown in \Cref{fig:RCS_example}.
It consists of frames, each frame comprising $n_f$ 5G NR sub-frames, hence lasting $T_f = n_f T_s$, where $T_s$ is the sub-frame time (referred to as slot time in the following).
Each frame comprises a \emph{contention time}, lasting one or more time slots, and \acp{SC} in all remaining time slots, where \ac{SC} definition and structure are as in sidelink.
The number of slot times devoted to contention time depends on the selected numerology.
In the case of $\mu = 0$ (reference numerology), it is assumed that contention time reduces to its minimum, one time slot.
Hence, the remaining $n_f-1$ slots of a frame are used to carry \acp{SC}.
If $n_{\text{SC}}$ \acp{SC} are configured in each time slot, the overall number of \acp{SC} in one frame is given by $K = (n_f - 1) n_{\text{SC}}$.
It is then possible to define a load factor $a$ as follows:
\begin{equation}
\label{eq:loadcoeff}
a = \frac{ n \, T_f }{ K \, T_g }
\end{equation}
Note that, if $a > 1$, the system is structurally overloaded, i.e., there are not enough \acp{SC} for all generated messages.

A delay budget is associated to each message.
If the time elapsing since when the message was generated exceeds the delay budget and the message has not been transmitted yet, the message is discarded, without being transmitted.
In the following, we assume that the delay budget is the same for all messages, denoted with $T_{\text{DB}}$.
To guarantee that at least one contention is possible irrespective of the generation time of the message, it must be $T_{\text{DB}} \ge 2 T_f$.
On the other hand, there is no point in setting $T_{\text{DB}} > T_g$, given that a new update message is generated every time period $T_g$.
Hence, we assume $2 T_f \le T_{\text{DB}} \le T_g$ (holding for $T_g > 2 T_f$).

%

\begin{table}[t]
\centering
\footnotesize
\setlength{\tabcolsep}{3pt}
\caption{Used symbols and corresponding definitions.}
\begin{tabular}{ll}
\hline
\textbf{Parameter} & \textbf{Definition} \\
\hline
$n$ & Number of nodes \\
$T_g$ & Message generation interval \\
$T_s$ & Slot duration \\
$T_f$ & Frame duration, $T_f = n_f T_s$ \\
$n_f$ & Number of slots per frame \\
$n_{\text{SC}}$ & Number of \acp{SC} per slot \\
$K$ & Number of \acp{SC} per frame, $K = (n_f-1)n_{\text{SC}}$ \\
$a$ & System load, given by $a = (n/T_g)/(K/T_f)$  \\
$M$ & Total number of sub-carriers used during contention \\
$J$ & Number of contention groups (divisor of $K$) \\
$w$ & Number of \acp{SC} associated with one contention group \\
$m$ & Sub-carriers per group, $m = \lfloor M/J \rfloor$ \\
$\delta$  & Duration of one contention round  \\
$T_{\text{DB}}$ & Packet delay budget \\
\hline
\end{tabular}
\label{tab:notation}
\vspace{-0.5cm}
\end{table}

\subsection{\ac{RCS} algorithm}
\label{subsec:ReCoalgo}

Let us follow the operation of a backlogged node, first introducing the baseline contention procedure (single-win).
Generalizations to multiple-win are then introduced.
Used notation is listed in \Cref{tab:notation}.

\subsubsection{Basic \ac{RCS} algorithm}
As soon as a new message is available, say at time $t_0$, the backlogged node waits until the first occurrence of the contention time after $t_0$.
Then it starts the contention procedure for that message.
Since there is one contention time per frame, the node must wait at most for a time $T_f - T_s$, where we account for a contention time equal to one slot time.
A set of sub-carriers is designated out of all sub-carriers in the sidelink channel.\footnote{We will see in \Cref{sec:experiment} that reliable detection of sub-carriers requires a frequency separation between sub-carriers used in the contention procedure.}
Let $M$ be the overall number of sub-carriers used in the contention procedure.
Since there are $K$ \acp{SC} to be assigned in one frame, sub-carriers are organized into $K$ subsets, each comprising $m$ sub-carriers.
Sub-carrier subset $j$ is associated to the $j$-th \ac{SC} of the frame, $j = 1,\dots,K$.\footnote{\acp{SC} belonging to one frame can always be numbered in a non-ambiguous order.}
A backlogged node selects one sub-carrier subset at random and contends for the \ac{SC} associated with that subset.

The contention time is organized into $r$ consecutive rounds.
At the beginning of the first round, the node selects at random one out of the $m$ sub-carriers of the chosen subset.
Let $f_1,f_2,\dots,f_m$ be the frequencies of the $m$ sub-carriers, labeled in increasing order, i.e., so that $f_i < f_j$ if $i < j$.
Let $f_x$ be the frequency picked at random by the considered node.
Each contending node transmits its selected sub-carrier for the whole round and \emph{simultaneously} the node listens to the frequency band comprising the selected sub-carrier group, to identify sub-carriers transmitted by other nodes.
If the considered node detects a frequency $f_y < f_x$, then the node deems itself as losing the contention and leaves the current contention.
It will take part in the next contention (after one frame time), if the packet delay budget allows this additional delay.
If instead the considered node does not detect any frequency less than its own selected frequency $f_x$, then it will deem itself as a winner, and it will move to the next round.
This contention procedure is repeated the same in every round, each time independently selecting a sub-carrier at random from the considered subset.
Nodes surviving up to the last round included (the $r$-th round) are the final winners.
Winners of sub-carrier subset $j$ are entitled to use \ac{SC} $j$ in the frame where they have won the contention.
A collision will occur if and only if there is more than one winner.
The analysis in \Cref{subsec:ReCoanalysis} will show that the probability of collision decays exponentially fast with the number of rounds.

After having won the contention and having transmitted in the \ac{SC} associated with the selected sub-carrier subset, the node goes back to idle if it has no other pending message.
Otherwise, it starts the whole procedure all over again for the next scheduled message.

\subsubsection{Generalized \ac{RCS} algorithm}
Splitting sub-carriers into $K$ groups and letting contending nodes joining groups at random may lead to some of the groups being void, i.e., with no contending node.
Then, the associated \ac{SC} would go unused in that frame.
To mitigate this potential waste of resource, we generalize the scheme of the previous sub-section as follows.
Let us define $J \le K$ resource groups, where $J$ is chosen as a divisor of $K$.
Each resource group comprises $w = K/J$ distinct \acp{SC} of a frame.
The basic procedure defined in the previous subsection is the special case where $J = K$ and hence $w = 1$.
Since multiple \acp{SC} are associated with each group, the contention in a group defines multiple winners.
More in depth, exactly $w$ winners should be ideally identified.
The number of actual winners in a given contention may fall short of $w$, because too few nodes joined that group.
It may also exceed $w$, in case multiple nodes select the same winning sub-carriers (then, a collision event will occur for at least one of the \acp{SC} belonging to the group).

To define the generalized procedure, the whole set of $M$ sub-carriers used for contention is split into $J$ subsets, each comprising $m = \lfloor M/J \rfloor$ sub-carriers.
Assume a contending node selects a sub-carrier with frequency $f_x$.
The contending node wins if it detects \emph{less} than $w$ sub-carriers with frequency $f_y < f_x$.
On the contrary, if at least $w$ sub-carriers with frequency $f_y < f_x$ are detected, the node leaves the contention.
This generalized contention is referred to as ``multi-win''.

%
%

\subsubsection{Implementation of the \ac{RCS} algorithm}

The critical points to implement the \ac{RCS} algorithm in 5G \ac{NR} and towards 6G sidelink are as follows.
\begin{itemize}
  \item Simultaneous transmission and reception of sub-carriers must be feasible.
  \item Nodes must be synchronized so that boundaries of contention rounds are aligned.
\end{itemize}

As for the first point, note that \ac{RCS} does not require implementing a true full-duplex radio.
The requirement to implement \ac{RCS} is only that a node be able to transmit a tone at frequency $f_x$ and \emph{at the same time} detect tones in a frequency band between $f_1$ and $f_{x-1}$, where $f_x$ is the tone frequency selected by the node.
The feasibility of this mild form of duplex is proved in experiments described in \Cref{sec:experiment}.

As for the second point, the duration $\delta$ of a contention round should be one or a few symbol times (with numerology $\mu = 0$, a sub-frame lasts 1 ms and contains 14 OFDM symbols, hence one symbol lasts $\approx$ \SI{71}{\micro\second}).
We will see in \Cref{sec:experiment} that a round time of 3 symbol times is enough to guarantee reliable simultaneous sub-carrier transmission and detection.
Tight synchronization among nodes operating in sidelink should be guaranteed anyway, independently of \ac{RCS}, to guarantee correct detection of \ac{SC} control information and carried data.

\subsection{Analysis of \ac{RCS} collision probability}
\label{subsec:ReCoanalysis}

A model is presented to evaluate the collision probability of the basic \ac{RCS} contention procedure \cite{baiocchi2019good}.
The model holds under the assumption that each contending node detects sub-carriers transmitted by all other contending nodes.

Let $n$ be the number of nodes backlogged at the beginning of the contention phase.
Let $q_i$ denote the probability that a node picks frequency $f_i$, $i=1,\dots,m$. 
Let also $G_i=\sum_{j=i}^{m}{q_j}$ be the corresponding \ac{CCDF}.

We define a \ac{DTMC} $X(t)$, where $X(t)$ denotes the number of nodes contending in round $t = 1,\dots,r$.
$X(0) = n$ is the initial number of contending nodes.
The \ac{DTMC} evolves over the state space $\{1,\dots,n\}$.
The one-step transition probability matrix of the \ac{DTMC} is denoted with $\mathbf{P}$, with entries given by:
\begin{equation}
\label{eq:Pdiag}
P_{k,h} = \begin{cases}
    \sum_{i=1}^{m-1}{ \binom{k}{h} q_i^hG_{i+1}^{k-h}}  & h=1,\dots,k-1  \\
    \sum_{i=1}^{m}{q_i^k}  &  h = k,  \\
    0  &  h = k+1,\dots,n,
\end{cases}
\end{equation}
for $k = 1,\dots,n$.

Let $\mathbf{p}(t)$ denote the state probability vector at time $t$, the $k$-th component of which is $p_k(t) = \mathcal{P}(X(t) = k), \, k = 1,\dots,n$.
At the outset of the contention we have $p_n(0) = 1$ and $p_k(0) = 0, \, k = 1,\dots,n-1$.
In the following, we choose a uniform probability distribution for sub-carrier selection in all rounds, i.e., we let $q_i = 1/m, \, i = 1,\dots,m$, for every round.
The corresponding matrix having entries as given in \Cref{eq:Pdiag} is denoted with $\mathbf{P}_u$, to emphasize that a uniform probability distribution has been adopted for sub-carrier selection.
Therefore, the probability distribution of the state of the \ac{DTMC} at time $t$ is $\mathbf{p}(t) = \mathbf{p}(0) \mathbf{P}_u^t$, for $t = 1,\dots,r$.
The probability of a successful contention, i.e., that a single node wins the last round, is $P_{\text{succ}} = \mathcal{P}( X(r) = 1 ) = p_1(r)$.
Correspondingly, the collision probability is $P_{\text{coll}} = \mathcal{P}( X(r) > 1 ) = 1-p_1(r)$.
Let $\mathbf{Q}_u$ denote the square matrix obtained by taking the last $n-1$ rows and columns of $\mathbf{P}_u$. 
The collision probability can be written as
\begin{equation}
\label{ }
P_{\text{coll}} = \mathbf{y} \, \mathbf{Q}_u^r \, \mathbf{e}
\end{equation}
where $\mathbf{y}$ is a row vector containing the last $n-1$ components of $\mathbf{p}(0)$, i.e., $y_{n-1} = 1$, $y_{k} = 0, \, k =1,\dots,n-2$, and $\mathbf{e}$ is a column vector of 1's.
The matrix $\mathbf{Q}_u$ is lower triangular, with diagonal elements given by $P_{u,kk}$ in \Cref{eq:Pdiag} for $k = 2,\dots,n$. 
Hence, its dominant eigenvalue is $\eta = Q_{u,11} = P_{u,22} = \sum_{i=1}^{m}{q_i^2} = 1/m$, where we have used the fact that $q_i = 1/m, \, \forall i$. 
Since $\mathbf{Q}_u$ is also a non-negative matrix, the right eigenvector  $\mathbf{u}$ associated to $\eta$ is positive. 
It is possible to find the closed form of $\mathbf{u}$, namely $\mathbf{u}^T = [2\,\, 3 \, \dots \, n]/2$, where the superscript $^T$ denotes transposition.
Since it is $\mathbf{e}^T = [1~1 \dots 1] \le [2~3 \dots n]/2 = \mathbf{u}^T$ and all vectors and matrices are non-negative, we have:
\begin{equation}
\label{eq:pcupperbound1}
P_{\text{coll}} = \mathbf{y} \mathbf{Q}_u^r \mathbf{e} \le \mathbf{y} \mathbf{Q}_u^r \mathbf{u} = \frac{ 1 }{ m^r } \mathbf{y} \mathbf{u} = \frac{ n }{ 2 m^r }
\end{equation}
In the end, we find the following upper bound on the collision probability for any value of the number of contending nodes, $n$:
\begin{equation}
\label{eq:pcupperbound2}
P_{\text{coll}} \le \min\left\{ 1 , \frac{ n }{ 2 m^r } \right\}
\end{equation}
The key feature of the upper bound is that it reveals that the collision probability decays exponentially as the number of rounds $r$ grows.
It also highlights the role of the number $m$ of sub-carriers used in the contention subset.
The analysis developed in this Section and the resulting upper bound hold as long as nodes can hear each other and sub-carrier detection is error-free.

\section{Simulations results}
\label{sec:sim}

This section presents the simulation-based evaluation of the proposed \ac{RCS} scheme. 
\cref{subsec:metrics_sim} defines the metrics used to assess both reliability and timeliness of packet delivery. 
The simulation setup is described in \cref{subsec:sim_setup}. \cref{subsec:rcs_optimize} analyzes the impact of the number of contention groups on reliability and timeliness over a wide range of load conditions. 
Finally, \cref{subsec:DS_SPS_RCS_results} compares the performance of \ac{RCS} with the standardized sidelink multiple access schemes \ac{SPS} and \ac{DS}.

\subsection{Key performance metrics for simulation of \ac{RCS}}
\label{subsec:metrics_sim}

It is assumed that all nodes can hear one another.
As a consequence, sub-carrier signaling works as described in \Cref{subsec:ReCoanalysis}.
An all-or-nothing model is used for the communication channel.
Then, packet reception is successful except in the case of collision (more than one node transmitting on the same \ac{SC}).
Hence, under \ac{RCS} possible outcomes of packet P generated by node A are as follows: (i) node A is the unique winner of an \ac{SC} resource for packet P; then P will be received successfully by all other nodes; (ii) multiple nodes win the same \ac{SC} as A; then packet P incurs a collision event that disrupts its reception; (iii) packet P is dropped (after having lost all available contentions), because it exceeds its delay budget.
The considered reliability metrics are related to outcomes of a packet: \ac{PRR}, collision probability ($P_{\mathrm{coll}}(sim)$), and packet drop probability ($P_{\mathrm{drop}}$), with $PRR + P_{\mathrm{coll}} + P_{\mathrm{drop}} = 1$.
The \ac{PRR} is defined as the fraction of successfully delivered packets:
\begin{equation}
\label{eq:PRR}
\mathrm{PRR} = \frac{N_{\mathrm{singlewin}}}{N_{\mathrm{total}}}
\end{equation}
The collision probability is defined as:
\begin{equation}
\label{eq:coll_sim}
P_{\mathrm{coll}} (sim) = \frac{N_{\mathrm{multiplewin}}}{N_{\mathrm{total}}}
\end{equation}
The packet drop probability accounts for delay constraints and is defined as:
\begin{equation}
P_{\mathrm{drop}} = \frac{N_{\mathrm{drop}}}{N_{\mathrm{total}}},
\end{equation}
where a packet is discarded if its delay exceeds $T_{DB}$.
A packet gets delayed when it loses a contention, and it has to go to the next contention.
Since each packet is associated with a deadline, if the elapsed time from generation exceeds this limit, the packet is considered obsolete and is discarded, giving up on running more contention.

The timeliness of the system is assessed through \ac{PIR}~\cite{3GPP_TR_37.885}, mean \ac{AoI}, and the \ac{AoI} violation probability. 
The \ac{PIR}~\cite{3GPP_TR_37.885} quantifies the time interval between two consecutive successful receptions of packets belonging to the same application flow, thus capturing the effective update timing at the receiver.
Consider the local dynamic map maintained at node $j$, which aggregates information received from neighboring nodes. 
Let $t_{ij}(g)$ denote the reception time at node $j$ of the $g$-th successfully received packet from node $i$, with $g \geq 2$. 
\ac{AoI} at node $j$ for updates from node $i$ is defined as
\begin{equation}
A_{ij}(t) = t - t_{ij}(g-1), \quad t \in [t_{ij}(g-1),\, t_{ij}(g))
\end{equation}
The peak \ac{AoI}, which coincides with \ac{PIR}, corresponding to the maximum age just before a new reception, is therefore
\begin{equation}
Y_{ij}(g) = t_{ij}(g) - t_{ij}(g-1)
\end{equation}
Let $z_{ij}$ denote the number of successfully received updates from node $i$ to node $j$. 
The average \ac{PIR} at node $j$, accounting for all incoming flows $(i,j)$, is computed as a weighted average:
\begin{equation}
\mathrm{E}[\mathrm{PIR}_j] = \sum_{i \in \mathcal{N}_j} \frac{\sum_{g=1}^{z_{ij}} Y_{ij}(g)}{\sum_{g \in \mathcal{N}_j} z_{gj}} \,
\end{equation}
where $\mathcal{N}_j$ is the set of nodes from which at least $\Omega$ packets have been successfully received by node $j$ (with $\Omega=2$ in simulations). 
The network-level \ac{PIR} is obtained by averaging $\mathrm{E}[\mathrm{PIR}_j]$ over all nodes.

\ac{AoI} measures the freshness of the most recently received information and applies to a wide range of communication scenarios, including cooperative awareness, perception, and coordination.
The average \ac{AoI} for updates from node $i$ to node $j$ is defined as
\begin{equation}
\label{eq:aoifromitoj}
\mathrm{E}[\mathrm{AoI}_{ij}] = \lim_{T \to \infty} \frac{1}{T} \int_{0}^{T} A_{ij}(t)\, dt 
\;\approx\; 
\frac{\sum_{g=1}^{z_{ij}} \frac{1}{2} Y_{ij}^2(g)}{\sum_{g=1}^{z_{ij}} Y_{ij}(g)}
\end{equation}

The relation between \ac{AoI} and \ac{PIR} can be expressed as
\begin{equation}
\label{eq:aoi_short}
\mathrm{E}[\mathrm{AoI}] = \frac{\mathrm{E}[\mathrm{PIR}^2]}{2 \cdot \mathrm{E}[\mathrm{PIR}]},
\end{equation}
showing that \ac{AoI} depends on both the mean and the variability of inter-reception intervals.
Based on inter-reception intervals, the probability of \ac{AoI} violation with respect to a threshold $A_{\text{th}}$ is defined for each node pair $(i,j)$ as
\begin{equation}
\label{eq:aoi-th}
V_{ij}(A_{\text{th}}) = \frac{\sum_{g=2}^{z_{ij}} \max\left\{0,\, Y_{ij}(g) - A_{\text{th}} \right\}}{\sum_{g=2}^{m_{ij}} Y_{ij}(g)}
\end{equation}
The global \ac{AoI} violation probability is obtained by averaging $V_{ij}$ over node pairs $(i,j)$.

\subsection{Simulation setup}
\label{subsec:sim_setup}

Simulations consider two numerologies, $\mu=0$ and $\mu=1$, corresponding to sub-carrier spacings of $15$~kHz and $30$~kHz, respectively. 
The only frame parameters kept fixed across numerologies are the bandwidth part, $BWP=20$~MHz, and the frame duration, $T_f=10$~ms. 
The time slot duration depends on the numerology, namely $T_s=1$~ms for $\mu=0$ and $T_s=0.5$~ms for $\mu=1$.
The frame partition into contention time and \acp{SC} is adapted accordingly. 

For $\mu=0$, the contention time occupies one slot, corresponding to $1$~ms, over the full $20$~MHz bandwidth. 
The remaining nine slots per frame host \acp{SC}, i.e., the resource to be allocated.
Each \ac{SC} is composed of $25$ \acp{RB}. 
With $4$ \acp{SC} per slot, $36$ \acp{SC} per frame are available for allocation.

For $\mu=1$, the full available bandwidth for two slot times is devoted to contention, so that contention time still amounts to 1 ms.
The data part of the frame, therefore, consists of the remaining $18$ slot times. 
Assigning still 25 \acp{RB} to one \ac{SC}, each slot accommodates $2$ \acp{SC}.
Then, one frame provides $36$ \acp{SC} for allocation, as in the case of $\mu = 0$. 
This design ensures that the two numerologies use the same frame duration, with the same contention time and the same amount of assignable \acp{SC}, thus enabling a fair comparison in the subsequent experiments.

Given the numerology dependent SCS and the overall bandwidth, $1200$ and $600$ sub-carriers are available for $\mu=0$ and $\mu=1$, respectively.
As we will see in \Cref{sec:experiment}, reliable sub-carrier detection is attained by using one sub-carrier out of three consecutive ones.
Then, $400$ and $200$ sub-carriers are available for contention in case of $\mu=0$ and $\mu=1$, respectively. 

Message generation time is set to $T_g = 100$~ms, delay budget to $T_{\mathrm{DB}} = 60$~ms, and AoI violation threshold to three different value: $100$~ms, $200$~ms and $500$~ms.
There are $14$ \ac{OFDM} symbols per slot.
The channel load coefficient $a$ (see \Cref{eq:loadcoeff}) is used as the independent variable.
Simulation time is $T_{\mathrm{sim}}=50$~s.

As for \ac{RCS} configuration, the number of contention rounds is fixed to $r = 3$, thus assigning three OFDM symbol times to each round (about \SI{213}{\micro\second}), with one symbol time gap between subsequent rounds (according to results obtained in \cref{sec:experiment}). 
In contrast to \ac{DS} and \ac{SPS}, only $36$ \acp{SC} are available for allocation per frame, since one $1$~ms portion of the frame is reserved for contention. 
The number of contention groups takes values in the set $\{1,2,3,6,9,18,36\}$, which corresponds to $\{36,18,12,6,4,2,1\}$ \acp{SC} associated with each group. 
Accordingly, the number of sub-carriers per group is $\{400,200,133,66,44,22,11\}$ for $\mu=0$ and $\{200,100,66,33,22,11,5\}$ for $\mu=1$.

As for \ac{SPS} configuration, the resource reservation interval is set to $\mathrm{RRI} = 100$~ms, the persistence probability is set to two values $P_{pers}=0$ and $P_{pers}=0.8$, and the \ac{RC} is uniformly drawn from the range $[5,15]$. 
In this case, all $40$ \acp{SC} in each frame are available for allocation. 
For a more detailed description of the operation of DS and SPS, the reader is referred to \cite{3GPP_TS_37.985,3GPP_TS_38.321,rolich2025rethink,lusvarghi2023comparative,garcia2021tutorial}.

\ac{DS} and \ac{SPS} are implemented in accordance with the model presented in~\cite{rolich2023,ROLICH2024}. 
In the \ac{SPS} implementation, we additionally incorporate the delay budget.
$T_{DB}$.
If, during the sensing and resource selection phase, insufficient resources are available for reservation, the packet is considered dropped. 
This does not occur in the \ac{DS} case, as transmission is performed in a one-shot manner. 
Another important modification of the model is that the channel load can exceed one, i.e., $N > K$. 
It is also assumed that all nodes are within mutual communication range, and that each message requires a single \ac{SC} for successful transmission, meaning it fits within one transport block.

\Cref{tab:parametervalues} summarizes the value of all relevant parameters.

\begin{table}[t]
\centering
\footnotesize
\setlength{\tabcolsep}{3pt}
\caption{Simulation parameters under different numerologies}
\begin{tabular}{lcc}
\hline
\textbf{Unified parameter} & \textbf{$\mu=0$} & \textbf{$\mu=1$} \\
\hline
\ac{SCS} & 15 kHz & 30 kHz \\
Slot duration ($T_s$) & 1 ms & 0.5 ms \\
Total \acp{RB} & 106 & 51 \\
Used sub-carriers ($M$) &400 & 200 \\
\acp{SC} per slot ($n_{SC}$) & 4 & 2 \\
Message generation time ($T_g$) & \multicolumn{2}{c}{100 ms} \\
Frame duration ($T_f$) & \multicolumn{2}{c}{10 ms} \\
Bandwidth ($BWP$) & \multicolumn{2}{c}{20 MHz} \\
\acp{RB} per \ac{SC} & \multicolumn{2}{c}{25} \\
OFDM symbols per slot & \multicolumn{2}{c}{14} \\
Time Delay Budget ($T_{DB}$) & \multicolumn{2}{c}{60 ms} \\
Load coefficient ($a$) & \multicolumn{2}{c}{[0.3; 1.1]} \\
Simulation time ($T_{sim}$) & \multicolumn{2}{c}{50 s} \\
AoI violation threshold ($A_{th}$) & \multicolumn{2}{c}{100, 200, 500 ms} \\

\hline
\textbf{DS parameter} & \textbf{$\mu=0$} & \textbf{$\mu=1$} \\
\hline
Number of \acp{SC} for allocation in frame ($K$)& \multicolumn{2}{c}{40} \\
Persistence probability ($P_{pers}$) & \multicolumn{2}{c}{0} \\
\ac{RC} & \multicolumn{2}{c}{1} \\

\hline
\textbf{SPS parameter} & \textbf{$\mu=0$} & \textbf{$\mu=1$} \\
\hline
\ac{RRI} & \multicolumn{2}{c}{100 ms} \\
Number of \acp{SC} for allocation in frame ($K$) & \multicolumn{2}{c}{40} \\
Persistence probability ($P_{pers}$) & \multicolumn{2}{c}{0, 0.8} \\
\ac{RC} & \multicolumn{2}{c}{[5; 15]} \\

\hline
\textbf{RCS parameter} & \textbf{$\mu=0$} & \textbf{$\mu=1$} \\
\hline
Max. number of contention rounds ($r$) & \multicolumn{2}{c}{3} \\
Number of \acp{SC} for allocation in frame ($K$) & \multicolumn{2}{c}{36} \\
Number of contention groups ($J$) & \multicolumn{2}{c}{1, 2, 3, 6, 9, 18, 36} \\
\acp{SC} per contention group ($w$) & \multicolumn{2}{c}{36, 18, 12, 6, 4, 2, 1} \\
Sub-carriers per group ($m_{J}$) & 400,200,133, & 200,100,66, \\
                      & 66,44,22,11 & 33,22,11,5 \\
\hline
\vspace{-0.7cm}
\end{tabular}
\label{tab:parametervalues}
\end{table}


\subsection{Tuning of \ac{RCS} parameters}
\label{subsec:rcs_optimize}

In this section, we compare different settings of \ac{RCS} parameters, focusing on different numbers of contention groups $J$ under various numerologies. 
All metrics presented are functions of the load $a$. 

\begin{figure}[t]
    \centering
    \subfloat[$\mu = 0$]{\label{fig:kpot_Psucc_mu0} 
    \includegraphics[width=.45\columnwidth]{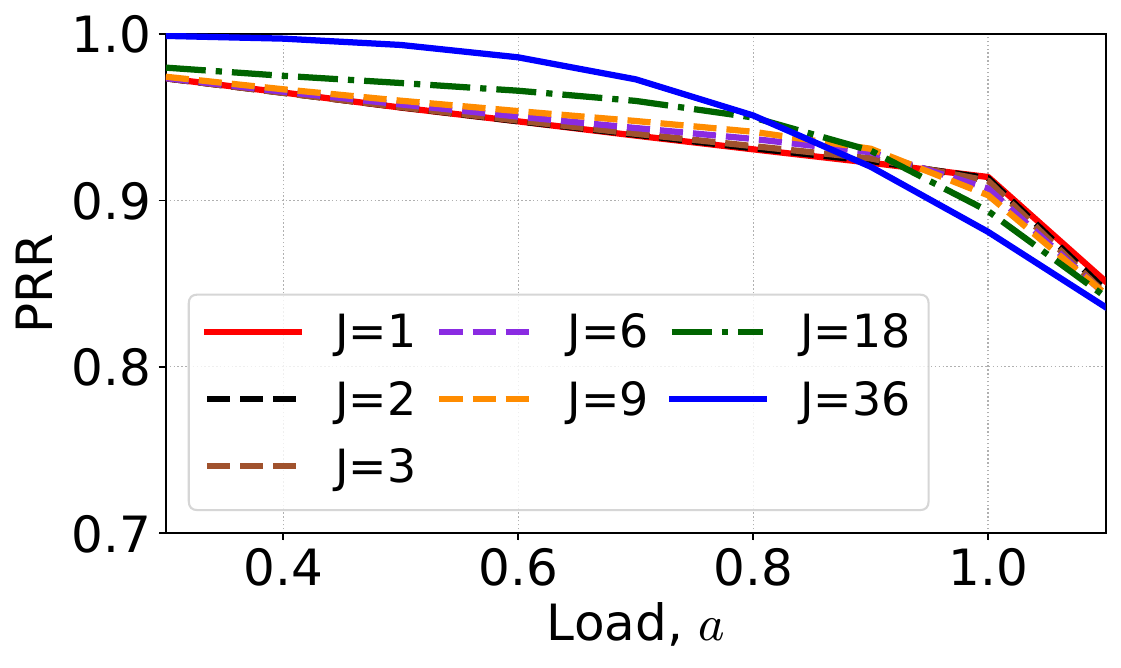}}%
    \subfloat[$\mu = 1$]{\label{fig:kpot_Psucc_mu1} 
    \includegraphics[width=.45\columnwidth]{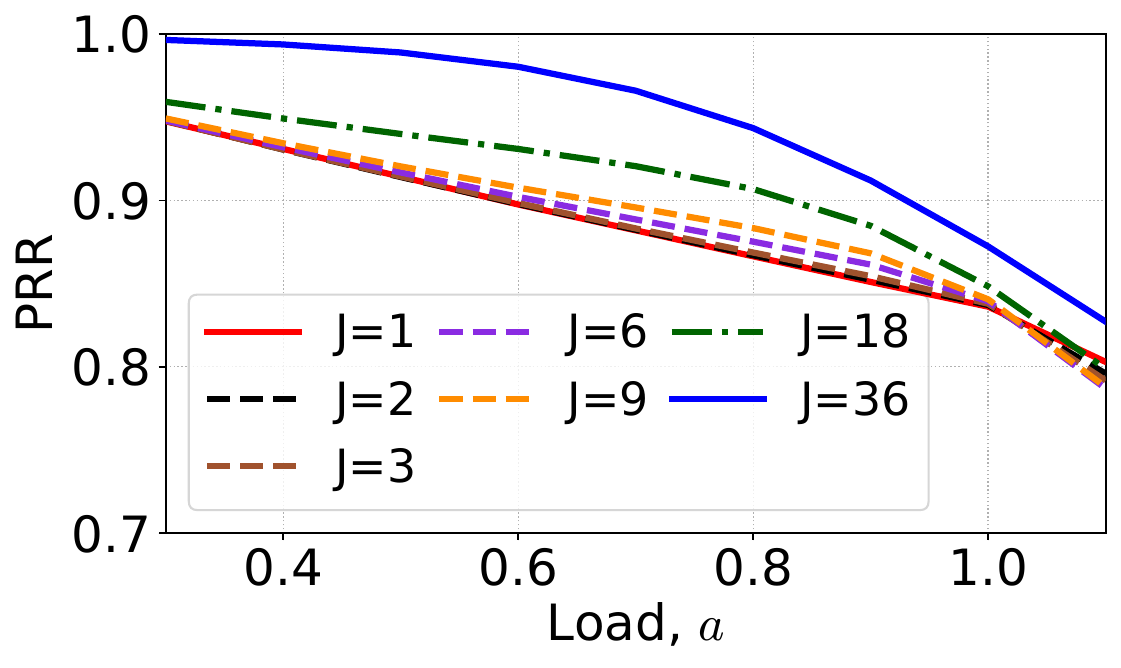}}%
    \caption{\ac{PRR} vs. load for different numbers of contention groups $J$.}
    \vspace{-0.5cm}
    \label{fig:kopt_Psucc}
\end{figure}

\cref{fig:kopt_Psucc} shows the \ac{PRR} for different values of $J$. 
As the load increases, the \ac{PRR} decreases for both values of $\mu$. 
For $\mu = 0$, at a load of $a = 0.9$, $J = 36$ is no longer optimal, and $J = 1$ becomes the best configuration in terms of successful packet delivery. 
In contrast, for $\mu = 1$, $J = 36$ consistently achieves the highest performance, while $J = 1$ yields the lowest. 
This occurs because the number of sub-carriers within a group for $\mu = 1$ is half that of the corresponding configuration at $\mu = 0$, leading to increased packet loss due to collisions or drops. 
For all other metrics, $J = 1$ and $J = 36$ represent boundary conditions, with intermediate values of $J$ yielding performance between these extremes.

\begin{figure}[t!]
    \centering
    \subfloat[$\mu = 0$]{\label{fig:kpot_Pcoll_mu0} 
    \includegraphics[width=.45\columnwidth]{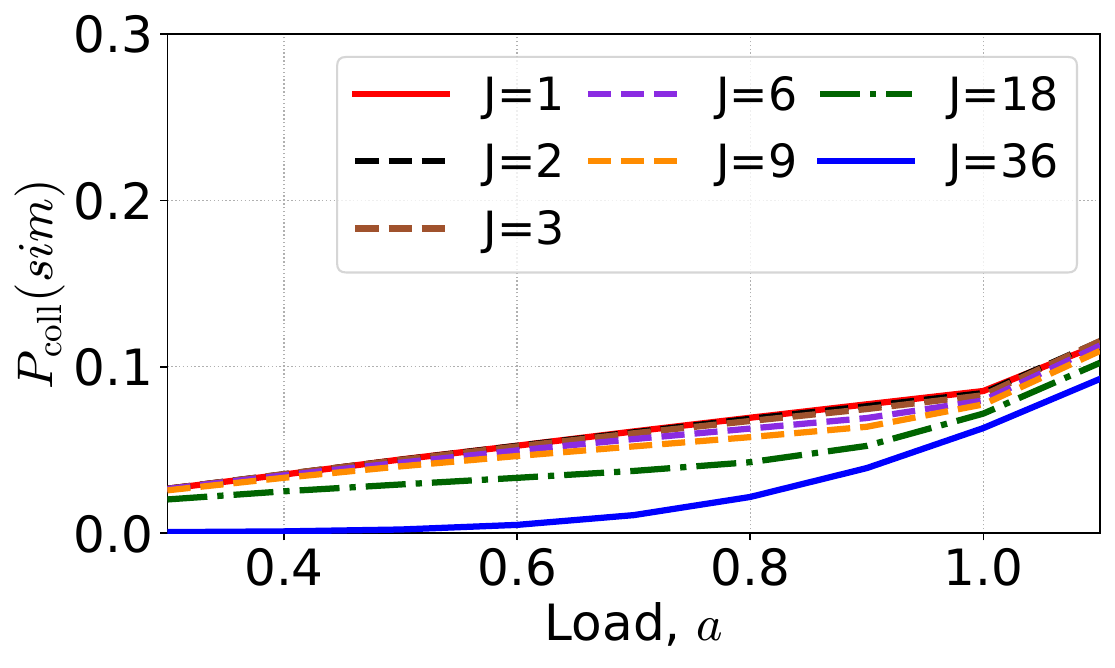}}%
    \subfloat[$\mu = 1$]{\label{fig:kpot_Pcoll_mu1} 
    \includegraphics[width=.45\columnwidth]{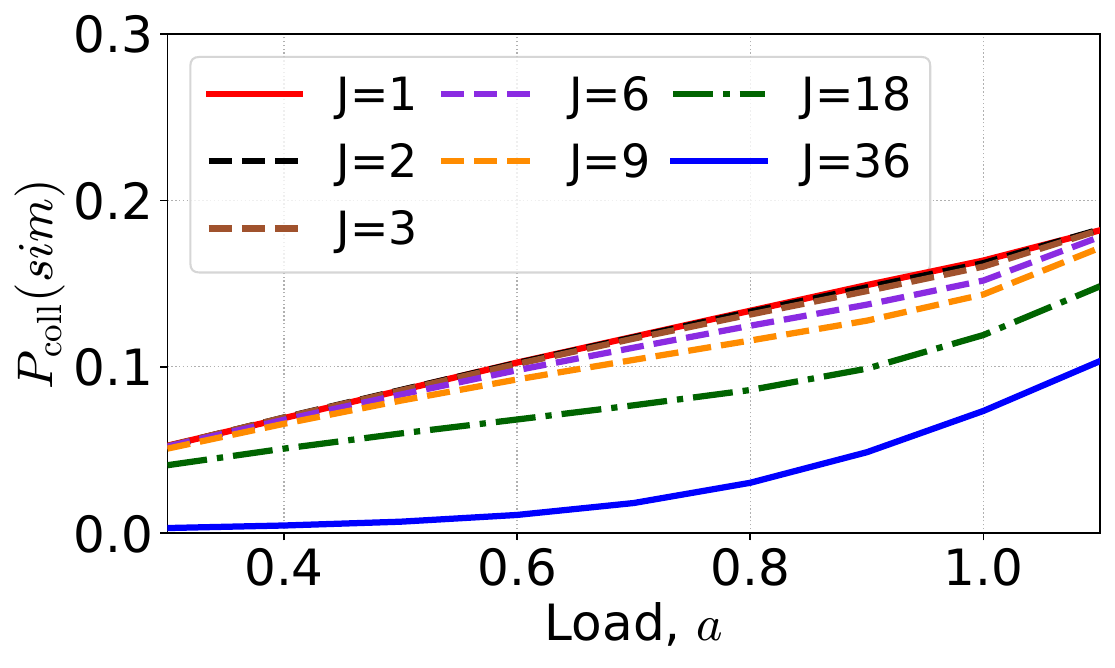}}%
    \caption{$P_{\mathrm{coll}}(sim)$ vs. load for different numbers of contention groups $J$.}
    \label{fig:kpot_Pcoll}
    \vspace{-0.5cm}
\end{figure}

\cref{fig:kpot_Pcoll} illustrates the collision probability, i.e., the probability that multiple winners emerge during the contention phase for a given \ac{SC}, causing simultaneous use of the same resources by multiple nodes. 
For $\mu = 0$, this probability is lower than for $\mu = 1$. 
For $J = 36$, performance remains nearly invariant with numerology, while for a smaller number of contention groups under $\mu = 1$, performance deteriorates. 
The collision probability increases with load in all cases.

\begin{figure}[t!]
    \centering
    \subfloat[$\mu = 0$]{\label{fig:kpot_Pdrop_mu0} 
    \includegraphics[width=.45\columnwidth]{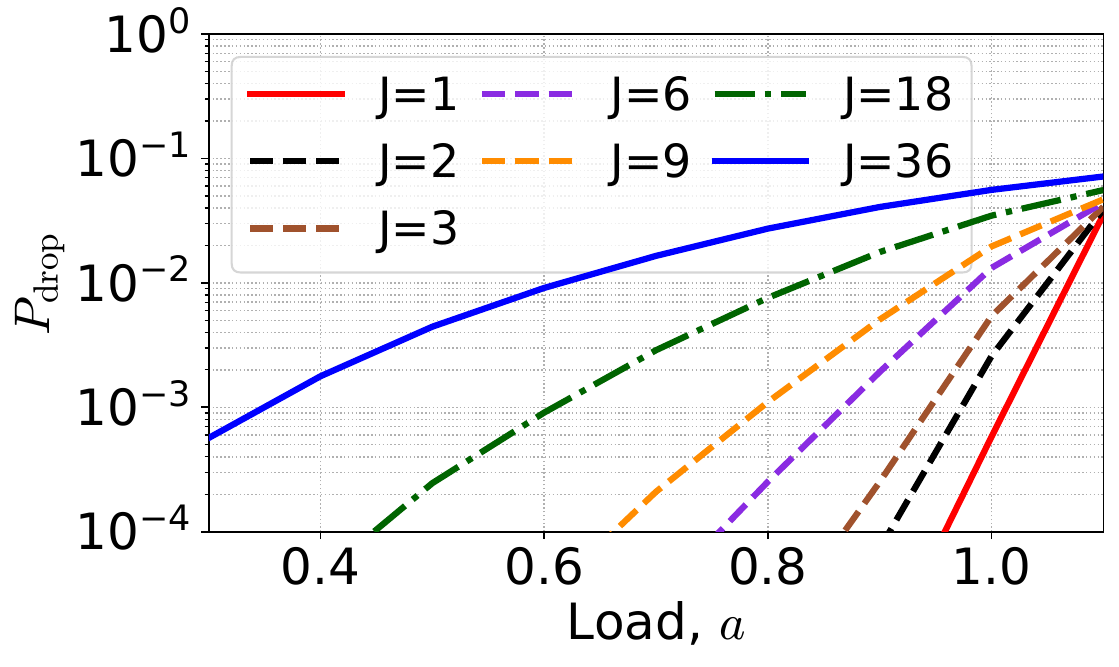}}%
    \subfloat[$\mu = 1$]{\label{fig:kpot_Pdrop_mu1} 
    \includegraphics[width=.45\columnwidth]{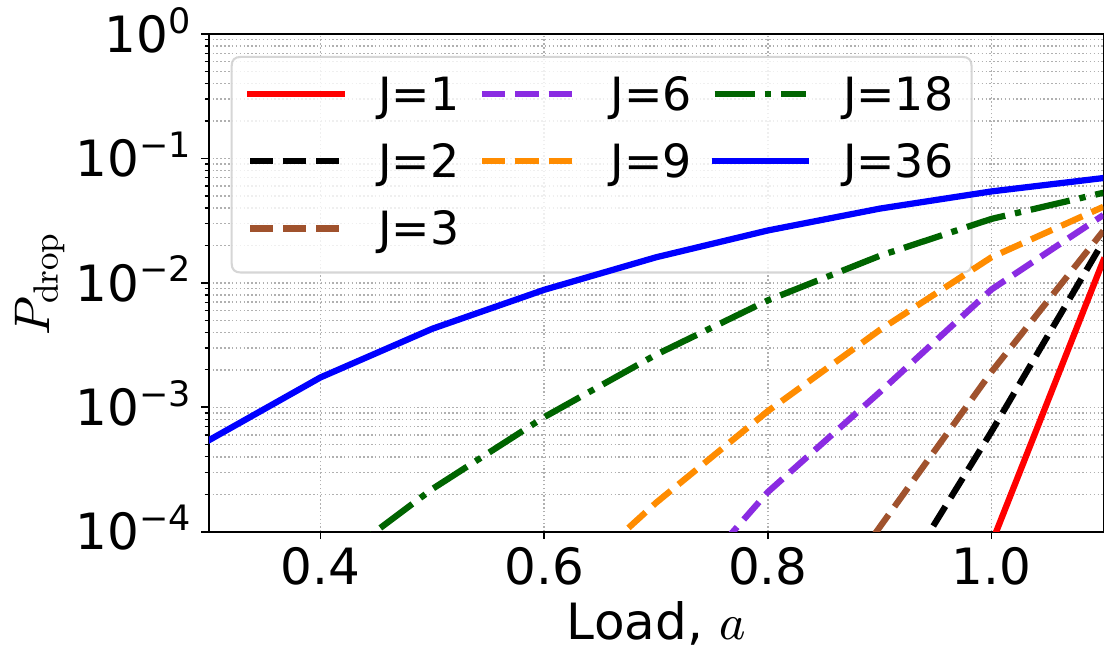}}%
    \caption{$P_{\mathrm{drop}}$ vs. load for different numbers of contention groups $J$.}
    \label{fig:kopt_Pdrop}
    \vspace{-0.5cm}
\end{figure}

\cref{fig:kopt_Pdrop} presents the packet drop probability. 
Here, $J = 36$ exhibits a substantially higher drop probability than $J = 1$. 
For $J = 1$ under low load, the drop probability is near zero and increases sharply only when $a > 0.9$. 
The figure uses a logarithmic scale, with maximum values remaining in the range [0.01, 0.1] even at $a = 1$. 
The drop probability is largely insensitive to numerology. 
The higher drop probability for $J = 36$ is due to the presence of only a single \ac{SC} in the group, resulting in more packets being discarded after the $T_{DB}$ during repeated contention, where competition is more intense.
Moreover, with $J = 36$ the probability that some group is deserted is not negligible, thus reducing the effectiveness of resource assignment.


\begin{figure}[t!]
    \centering
    \subfloat[$\mu = 0$]{\label{fig:kpot_PIR_mu0} 
    \includegraphics[width=.45\columnwidth]{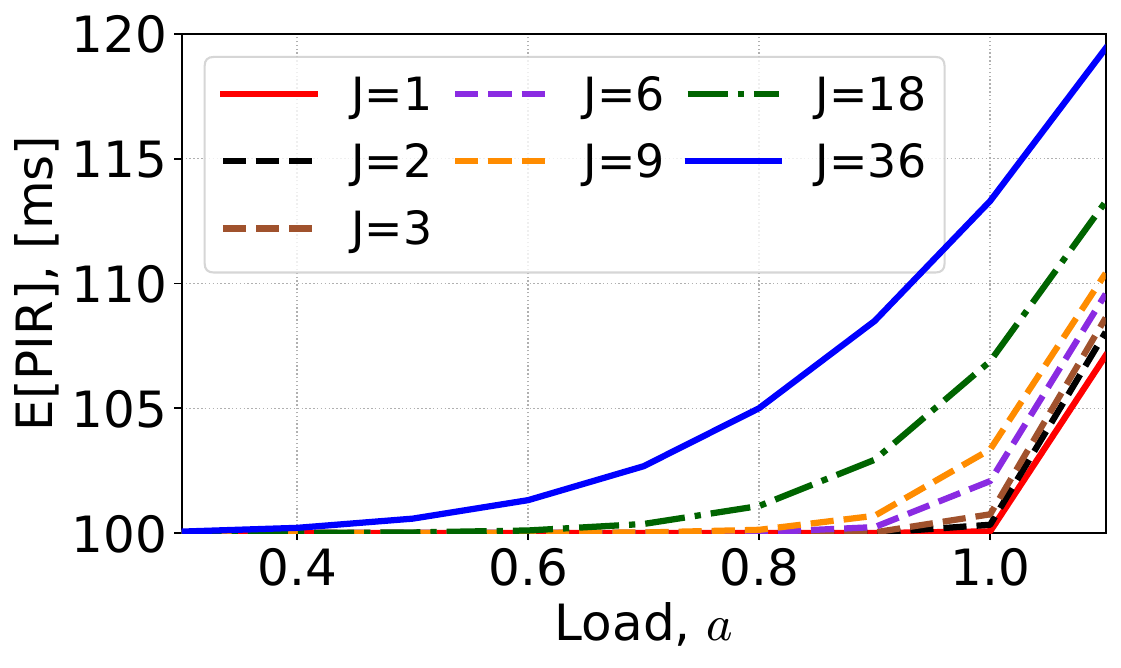}}%
    \subfloat[$\mu = 1$]{\label{fig:kpot_PIR_mu1} 
    \includegraphics[width=.45\columnwidth]{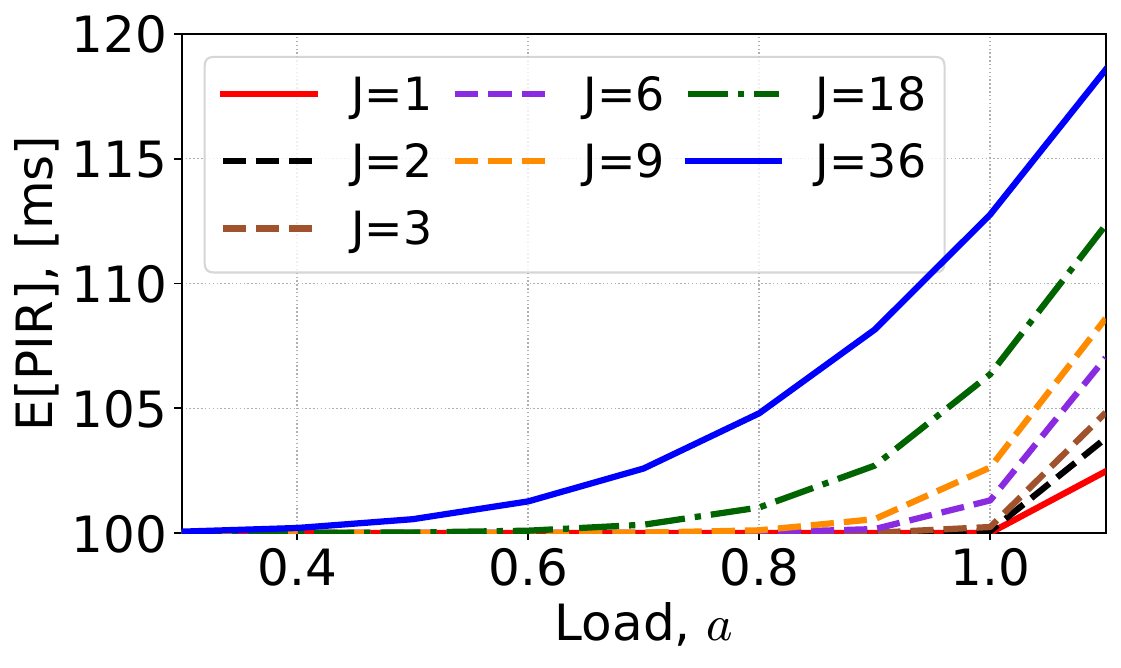}}%
    \caption{$E[PIR]$ vs. load for different numbers of contention groups $J$.}
    \label{fig:kopt_PIR}
    \vspace{-0.5cm}
\end{figure}

\begin{figure}[t!]
    \centering
    \subfloat[$\mu = 0$]{\label{fig:kpot_AoI_mu0} 
    \includegraphics[width=.45\columnwidth]{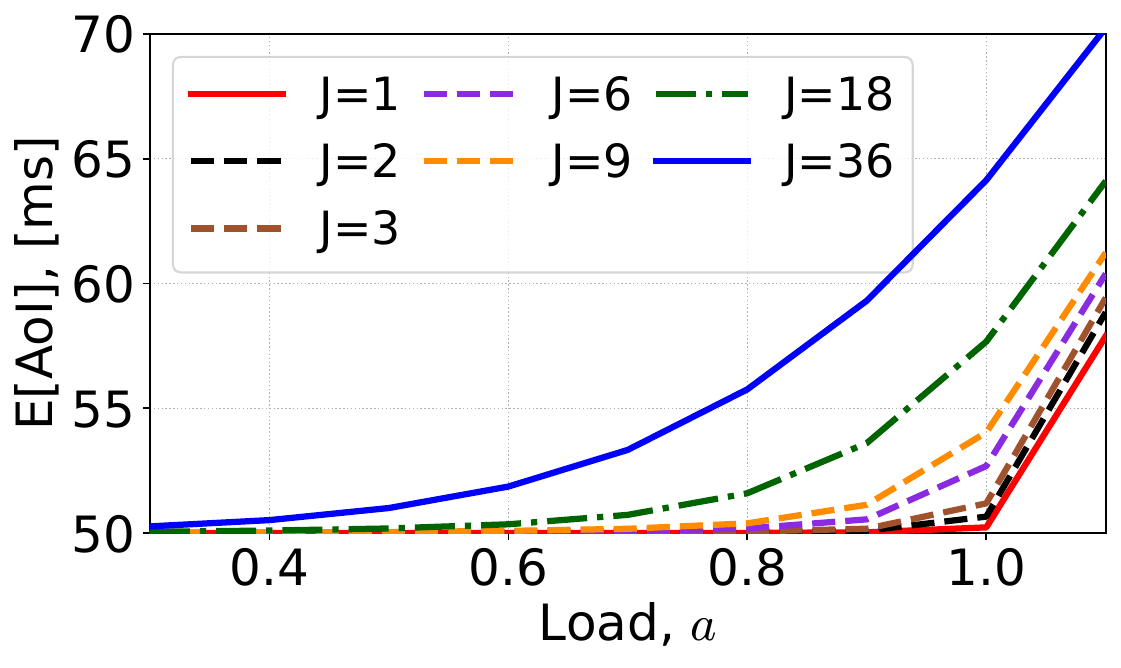}}%
    \subfloat[$\mu = 1$]{\label{fig:kpot_AoI_mu1} 
    \includegraphics[width=.45\columnwidth]{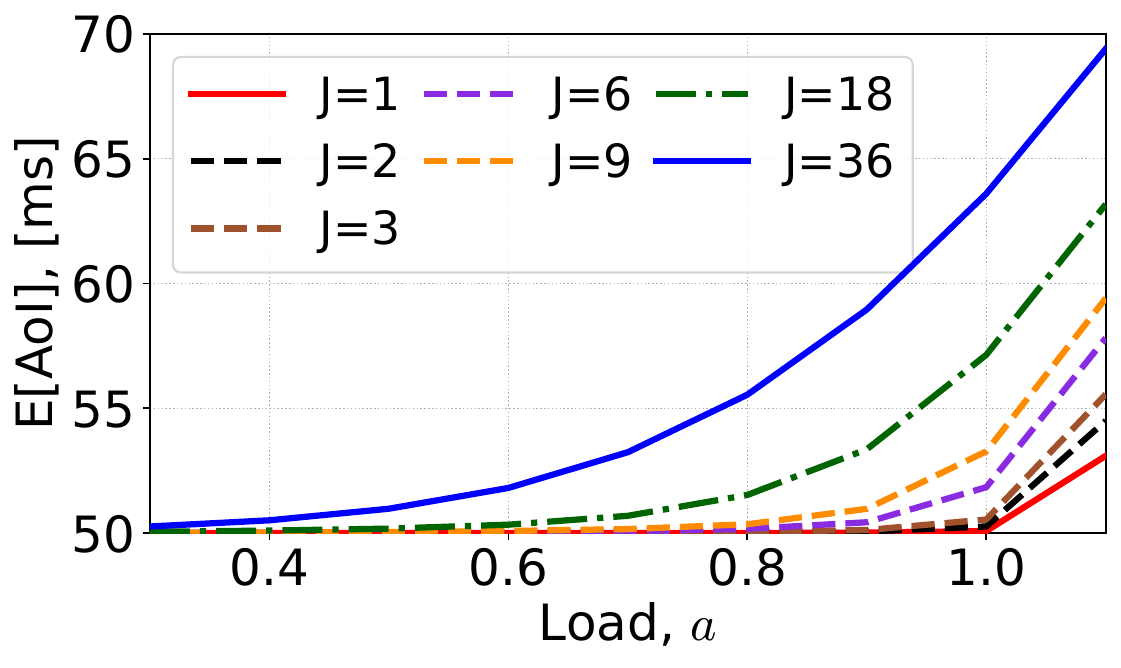}}%
    \caption{$E[AoI]$ vs. load for different numbers of contention groups $J$.}
    \label{fig:kopt_AoI}
    \vspace{-0.5cm}
\end{figure}

\cref{fig:kopt_PIR} and \cref{fig:kopt_AoI} show the mean \ac{PIR} and mean \ac{AoI}, respectively. 
Both metrics increase with load due to more frequent collisions and longer intervals between successful transmissions. 
For $\mu = 0$ and $\mu = 1$, $J = 1$ yields the best performance. 
The absolute differences in performance remain moderate, on the order of 15 ms under high load.

These results reveal two representative operating regimes for \ac{RCS}. 
At low load, $J = 36$ achieves the highest probability of successful reception and the lowest collision probability, as losses are dominated by collisions and a larger number of groups reduces contention effectively. 
As the load increases, this advantage diminishes, and performance across group configurations converges. 
In the high-load regime ($a \geq 0.9$), $J = 1$ becomes preferable, offering better timeliness and the highest successful transmission probability, while $J = 36$ becomes slightly less effective. 
Overall, \ac{RCS} maintains satisfactory performance, with collision probabilities on the order of $10^{-1}$ at the maximum considered load. 
In this regime, packet loss is increasingly influenced by drop events, as packets cannot always be delivered within the available time budget.

Therefore, in \cref{subsec:DS_SPS_RCS_results}, we consider only the boundary cases $J = 1$ and $J = 36$ for \ac{RCS} and restrict the analysis to $\mu = 1$, as \ac{DS} and \ac{SPS} performance is independent of $\mu$ with given channel configuration, while \ac{RCS} is evaluated in its worst-case \ac{PRR} configuration.

\subsection{Comparison with standardized sidelink multiple access algorithms}
\label{subsec:DS_SPS_RCS_results}

In this section, we compare the configurations \ac{RCS} with $J = 1$ and $J = 36$, \ac{DS}, and the standardized persistence probability values $0$ and $0.8$ for \ac{SPS}. We analyze the performance of the standardized resource selection algorithms and the proposed \ac{RCS} using the \ac{PRR}, mean {PIR}, and mean \ac{AoI}. Additionally, we examine the \ac{AoI} violation probability for three thresholds: 100 ms, 200 ms, and 500 ms. All metrics presented are functions of the load $a$. 

\begin{figure*}[t]
\centering\includegraphics[width=0.95\textwidth]{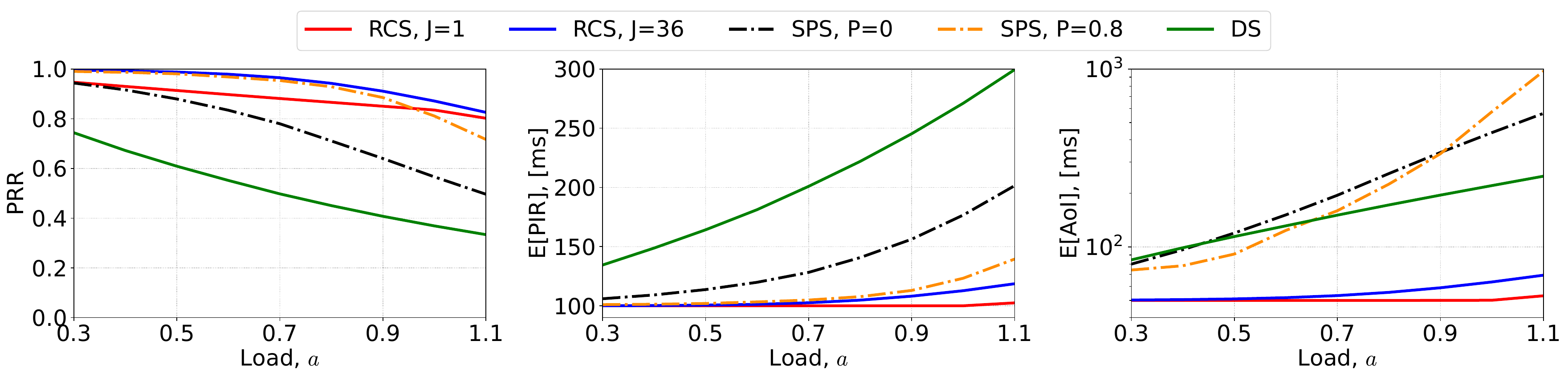}
\caption{Comparison of DS, SPS, and RCS performance in terms of PRR (left), $\mathbb{E}[\mathrm{PIR}]$ (center), and $\mathbb{E}[\mathrm{AoI}]$ (right).}
\label{fig:DS_SPS_RCS_metrics}
\vspace{-0.2cm}
\end{figure*}

\begin{figure*}[t]
\centering\includegraphics[width=0.95\textwidth]{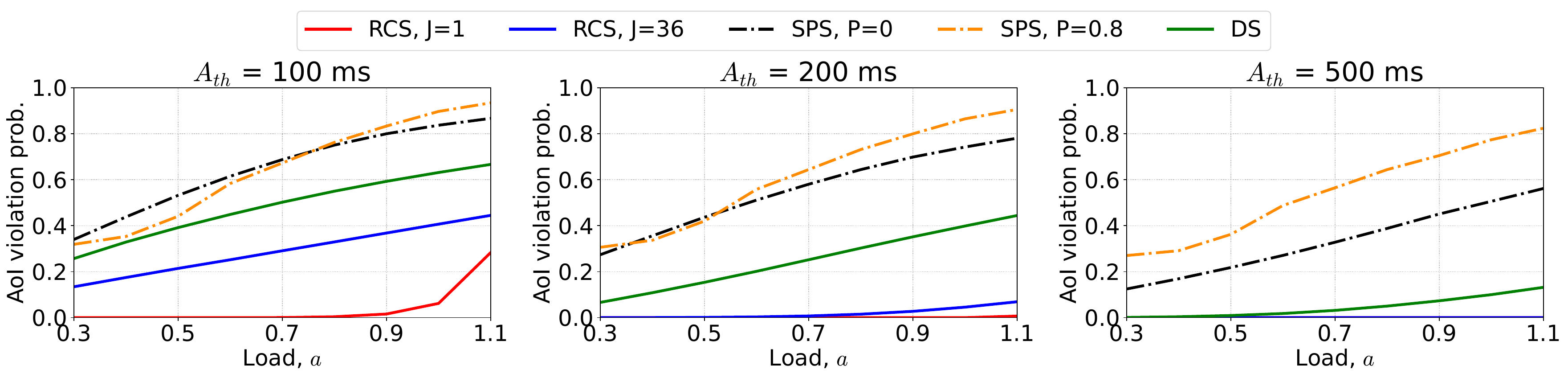}
\caption{Comparison of DS, SPS, and RCS in terms of AoI violation probability for different thresholds $A_{th}$: left ($A_{th}=100 \mathrm{ms}$), center ($A_{th}=200 \mathrm{ms}$), and right ($A_{th}=500 \mathrm{ms}$).}
\label{fig:violAoI_comp}
\vspace{-0.5cm}
\end{figure*}

\cref{fig:DS_SPS_RCS_metrics} presents three metrics: \ac{PRR} (left), mean \ac{PIR} (center), and mean \ac{AoI} (right, plotted on a logarithmic scale). The results show that the proposed \ac{RCS} can outperform the standardized \ac{SPS} and \ac{DS} under periodic network traffic.  
From the perspective of \ac{PRR}, the advantage of \ac{RCS} over \ac{SPS} is moderate, as the curves for different \ac{RCS} and \ac{SPS} configurations intersect. Nevertheless, \ac{RCS} consistently outperforms \ac{DS} and \ac{SPS} with a persistence probability of 0 across all load conditions and matches the \ac{PRR} of \ac{SPS} only when the persistence probability is 0.8.

The analysis of the mean \ac{PIR} reveals a clear trend: as the load increases, the average interval between successfully delivered packets grows. This increase is steeper for \ac{DS} and \ac{SPS}, particularly at high channel loads around $0.8$–$0.9$. At low loads, \ac{SPS} with a persistence configuration of 0.8 is comparable to \ac{RCS}, but under higher loads, \ac{RCS} clearly outperforms \ac{SPS}.

Considering the mean \ac{AoI} as a function of load, the proposed \ac{RCS} significantly surpasses all standardized resource selection algorithms. In \ac{DS}, collisions are frequent but random and uncorrelated, whereas \ac{SPS} experiences consecutive persistent collisions and burst losses, leading to a sharper increase in mean \ac{AoI}.

This behavior is further illustrated in \cref{fig:violAoI_comp}, which shows the \ac{AoI} violation probability. 
\ac{RCS} outperforms \ac{SPS} and \ac{DS} across all thresholds, including the stringent 100 ms threshold, demonstrating that \ac{RCS} effectively mitigates persistent collisions and bursty losses.

For \ac{SPS}, the violation probability changes little across thresholds, indicating that many \ac{AoI} samples exceed 500 ms, reflecting persistent collisions that block communication for extended periods.
In contrast, for \ac{DS} and \ac{RCS}, the violation probability decreases with increasing thresholds, indicating fewer extreme \ac{AoI} values, consistent with observations in~\cite{rolich2025rethink,rolich2025tradeoff}. This demonstrates more regular and reliable packet delivery. Notably, for the 200 ms threshold, \ac{RCS} exhibits extremely low violation probabilities, not exceeding 0.1 even under high load. For the 500 ms threshold, the probability approaches zero, showing that messages are delivered consistently even in the presence of collisions and drops, ensuring stable communication.

These results indicate that \ac{RCS} outperforms the standardized \ac{DS} and \ac{SPS} under the worst-case configuration ($\mu = 1$), providing high packet delivery reliability and regular, stable updates between nodes. This improvement arises from the complete elimination of the sensing and resource reservation procedure, which significantly reduces the probability and duration of persistent collisions.

The obtained results also indicate that a reliability metric, such as \ac{PRR}, for \ac{RCS} is fully correlated with timeliness metrics, such as \ac{AoI}. In contrast, for \ac{SPS}, this correlation is nontrivial and not immediately evident, due to the hidden issue of persistent collisions, including those that may last for extended periods.

\section{Experimental proof-of-concept}
\label{sec:experiment}
This section presents the experimental validation of the \ac{RCS} scheme. \cref{subsec:metrics_exp} defines metrics for reliability and feasibility. \cref{subsec:setup_exp} details the hardware, PHY, and synchronization setup. \cref{subsec:tuning_exp} examines key PHY and protocol parameters, and \cref{subsec:results_exp} compares experimental results with simulations, highlighting real-world non-idealities.

\subsection{Key performance metrics for experimental evaluation of \ac{RCS}}
\label{subsec:metrics_exp}
The considered reliability metrics are: probability of success ($P_{\mathrm{succ}}$), collision probability ($P_{\mathrm{coll}}(exp)$), no-winner probability ($P_{\mathrm{nowin}}$).
Unlike simulations, the experiment yields three possible outcomes: (i) success with a single winner; (ii) collision with multiple winners; and (iii) contention terminating without any winner.
Hence it is $P_{\mathrm{succ}} + P_{\mathrm{coll}}(exp) + P_{\mathrm{nowin}} = 1$ and the probability of packet loss is $P_{\mathrm{loss}} = P_{\mathrm{coll}}(exp) + P_{\mathrm{nowin}}$.

In the following calculations, $P_{\mathrm{succ}}$ is equivalent to the \ac{PRR} (\cref{eq:PRR}), while the collision probability defined in simulations, $P_{\mathrm{coll}}(\mathrm{sim})$ (\cref{eq:coll_sim}), corresponds directly to its experimental counterpart, $P_{\mathrm{coll}}(\mathrm{exp})$.

The no-winner probability, observed only in experiments, is defined as the fraction of contentions that end without selecting any winner, relative to the total number of contentions. It is given by:
\begin{equation}
P_{\mathrm{nowin}} = \frac{N_{\mathrm{nowin}}}{N_{\mathrm{total}}}
\end{equation}

To identify a suitable configuration, we evaluated the false positive probability ($P_{fp}$), false alarm probability ($P_{fa}$), and the detection probability ($P_d$). A false alarm occurs when a node detects a non-existent tone, potentially causing all nodes to withdraw and leaving the resource unassigned. A missed detection occurs when a transmitted tone is not detected, which may result in multiple nodes claiming the resource and causing a collision. The false positive probability is defined as the probability that a node detects a tone on a sub-carrier that immediately precedes the transmitted one, typically due to spectral leakage caused by the signal's shape.

\subsection{Experimental setup}
\label{subsec:setup_exp}

The experimental setup implements a wireless node prototype capable of simultaneous tone transmission and channel sensing, as required for frequency-domain \ac{RCS} operation \cite{baiocchi2019good}. The system is built on a \ac{SDR} platform based on Ettus USRP B200 devices, enabling flexible physical layer control. Signal processing and protocol logic are implemented in GNU Radio, allowing modular integration of the contention mechanism with the radio front-end. Synchronization across nodes is achieved via an Ettus CDA-2990 OctoClock, which provides a common frequency reference and aligned sampling time.

The \ac{SDR} implementation introduces practical impairments, including noise, interference, limited dynamic range, and oscillator instability. These effects require careful parameter tuning, particularly for guard band allocation and sub-carrier selection, to ensure reliable operation under FR1 sidelink specifications~\cite{3GPP_TS_38.311}. The adopted PHY configuration, summarized in Table~\ref{tab:experiment_parameters}, balances standard compliance with real-time processing constraints. The system operates with numerology $\mu=0$, corresponding to a \ac{SCS} of 15 kHz and a slot duration of 1 ms.

Following the \ac{RCS} principle, contention levels are mapped onto \ac{OFDM} sub-carriers. Due to computational constraints, the FFT size is fixed to $N_{FFT}=32$, from which a subset of 10 sub-carriers is selected to represent contention levels.

Each node transmits a single OFDM tone by activating one randomly selected sub-carrier while sensing the channel simultaneously. If a tone is detected at a lower frequency, the node withdraws; otherwise, it proceeds to the next contention round. A tone is considered valid if its power exceeds a threshold determined by the noise floor.

To emulate a contention with multiple stations, one of the available USRP devices was programmed to  represent more than a single node, allowing the total number of competing nodes $n$ to range between 3 and 10. At each round, it generates simultaneous OFDM tones over $m$ sub-carriers and executes the \ac{RCS} procedure. The evaluation is based on an extensive experimental campaign comprising more than $10^5$ contention cycles per configuration.

\begin{table}[t]
\centering
\footnotesize
\setlength{\tabcolsep}{3pt}
\caption{Experimental parameters}
\begin{tabular}{lc}
\hline
\textbf{Parameter} & \textbf{Value} \\
\hline
Central frequency ($f_c$) & 5 GHz \\
FFT size ($N_{FFT}$) & 32 samples \\
Numerology ($\mu$) & 0 \\
\ac{SCS} & 15 kHz \\
Slot duration ($T_s$) & 1 ms \\
Bandwidth ($BWP$) & 480 kHz \\
OFDM symbols per slot & 14 \\
Symbol duration & 71.428 ms \\
Cycle Prefix length & 3 samples \\
\hline
\end{tabular}
\label{tab:experiment_parameters}
\vspace{-0.5cm}
\end{table}

\subsection{Experimental parameters tuning}
\label{subsec:tuning_exp}


\begin{table}[b]
\vspace{-0.5cm}
\centering
\caption{Mapping of contention sub-carriers indices and frequency offsets with respect to the center frequency.}
\label{tab:sub-carrier_mapping}
\begin{tabular}{@{}cc|cc@{}}
\toprule
Index & Offset (kHz) & Index & Offset (kHz) \\ \midrule
-15         & -225         & +1          & +15          \\
-12         & -180         & +4          & +60          \\
-9          & -135         & +7          & +105         \\
-6          & -90          & +10         & +150         \\
-3          & -45          & +13         & +195         \\ \bottomrule
\end{tabular}
\end{table}

\begin{figure*}[t!]
    \centering
    \subfloat[$P_{fp}$ as a function of guard carriers number]{\label{fig:pfalse} 
    \includegraphics[width=.32\textwidth]{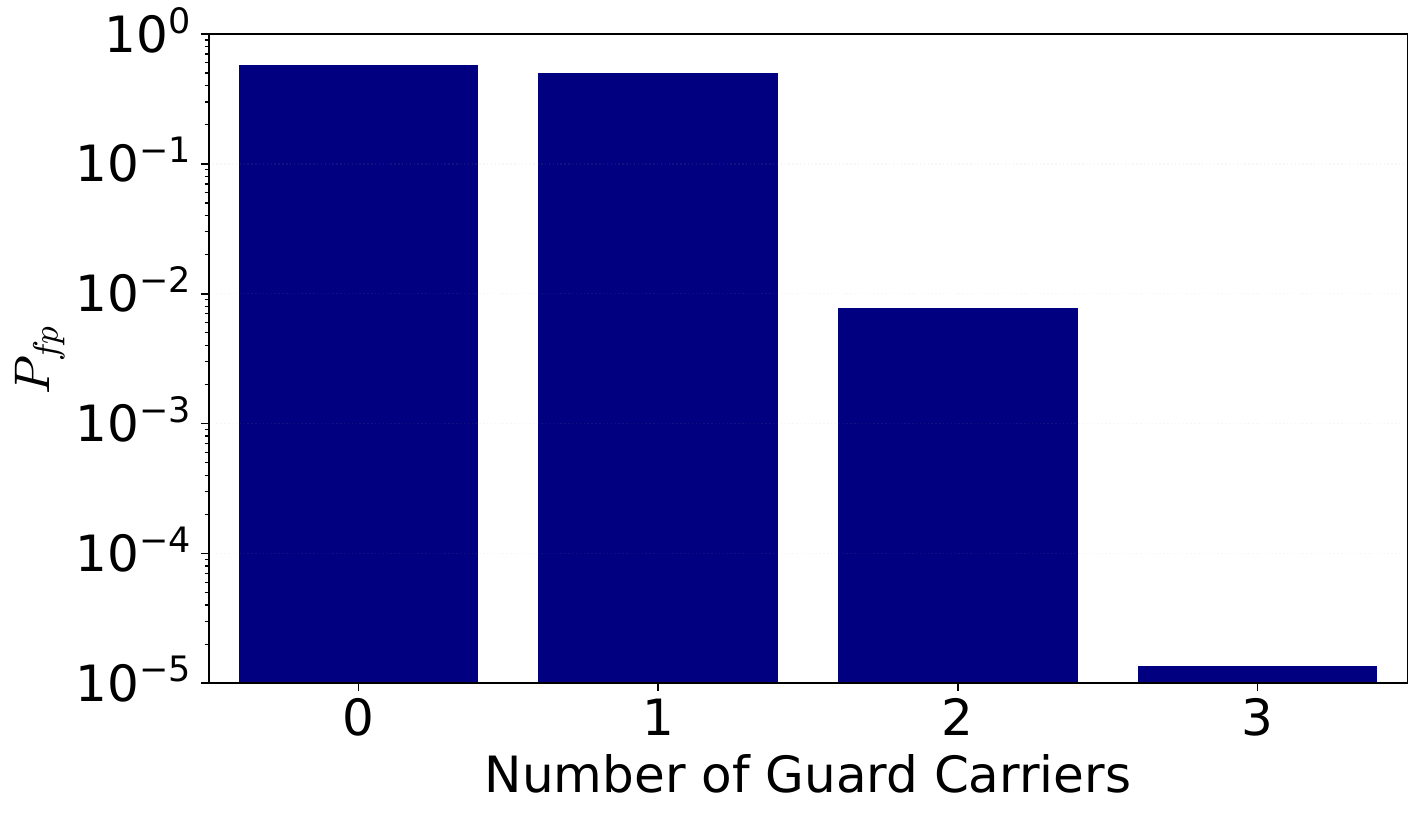}}%
    \subfloat[$P_{fa}$ as a function of round duration]{\label{fig:roundvspfa} 
    \includegraphics[width=.32\textwidth]{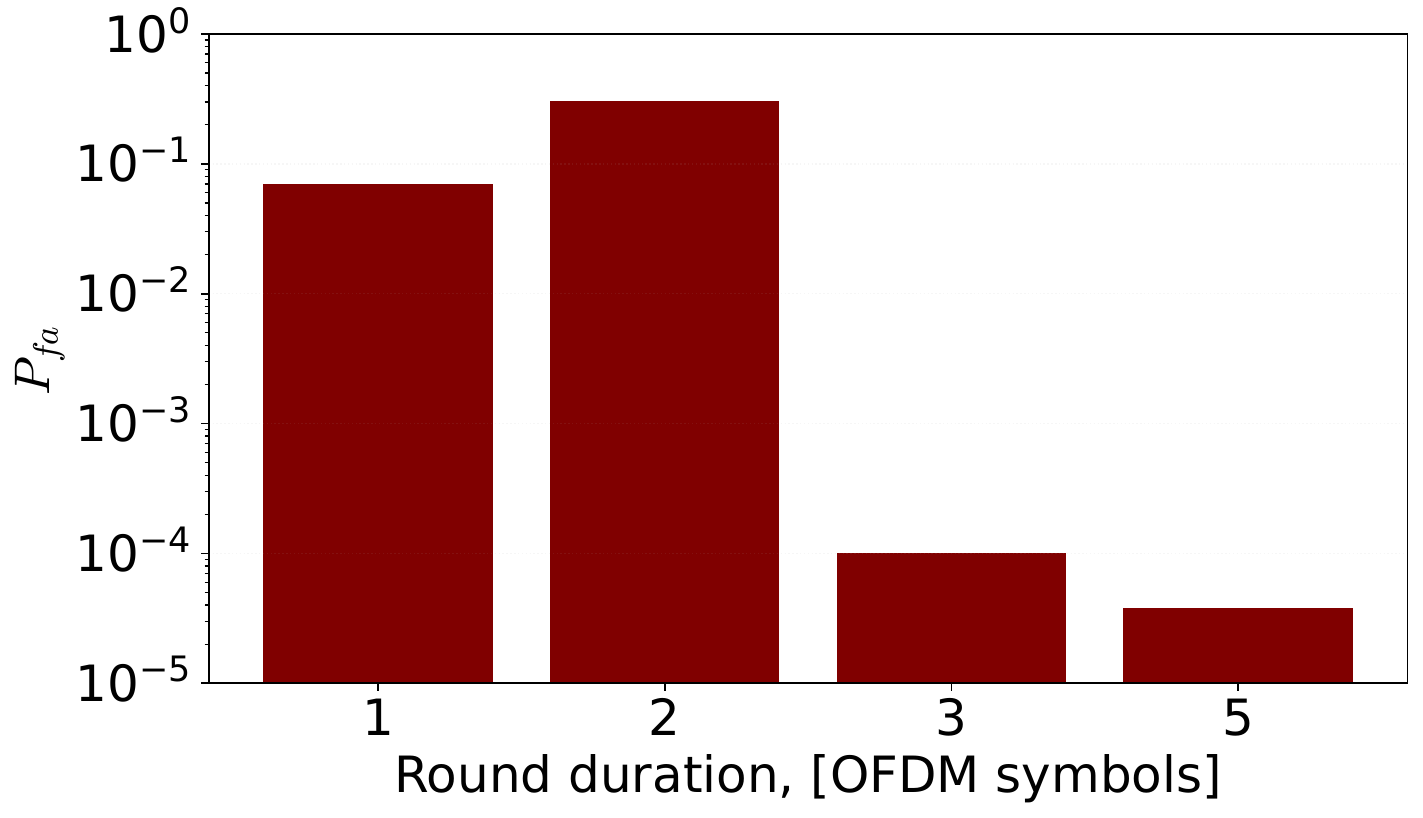}}%
    \subfloat[$P_{d}$ as a function of round duration]{\label{fig:roundvspd} 
    \includegraphics[width=.32\textwidth]{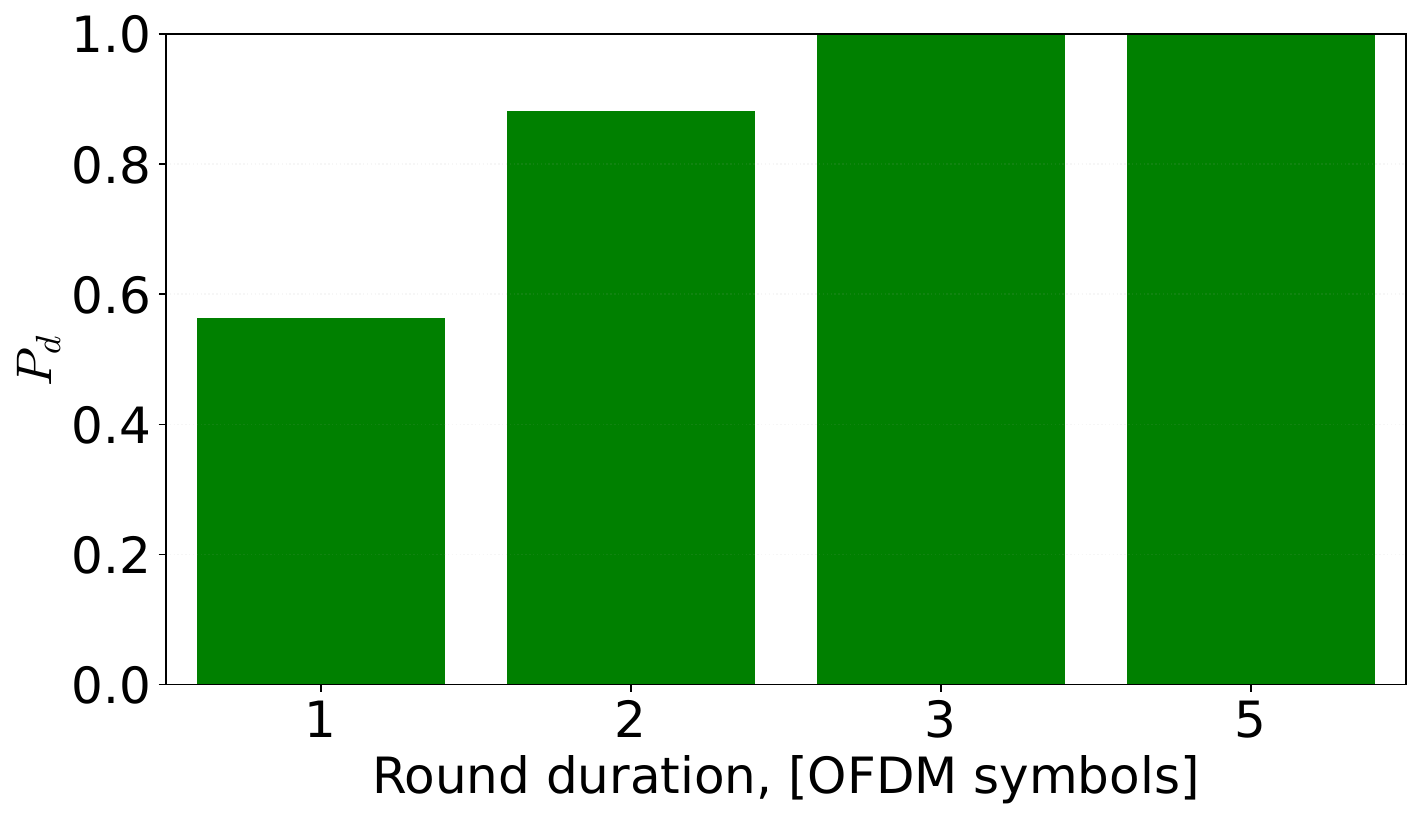}}%
    \caption{Results of experimental parameters tuning.}
    \label{fig:combined_metrics}
    \vspace{-0.5cm}
\end{figure*}

The sub-carrier selection (see \cref{tab:sub-carrier_mapping}) mitigates hardware non-idealities by introducing a spacing of 2 sub-carriers between contention levels, reducing inter-carrier interference, and excluding the DC and mirror frequencies to avoid sensing errors due to local oscillator leakage and image components. This spacing also limits false positives caused by spectral leakage into adjacent bins, which may lead nodes to prematurely withdraw from contention. Evaluation of the false positive probability for different guard spacings shows that using 2 guard sub-carriers achieves $P_{fp}$ of approximately $10^{-2}$ (see \cref{fig:pfalse}), which is acceptable for the experimental setup.

The round duration determines the trade-off between detection reliability and the number of rounds per slot: short durations degrade detection performance, while longer durations reduce the number of available rounds. Given the 14-symbol slot structure, with the first and last symbols reserved for PHY functions~\cite{garcia2021tutorial} and a guard symbol required between consecutive rounds, four time-slot configurations are feasible:
\begin{itemize}
    \item 6 rounds of 1 symbol each, with 6 guard symbols;
    \item 4 rounds of 2 symbols each, with 4 guard symbols;
    \item 3 rounds of 3 symbols each, with 3 guard symbols;
    \item 2 rounds of 5 symbols each, with 2 guard symbols.
\end{itemize}
The false alarm ($P_{fa}$) and detection ($P_d$) probabilities were evaluated for each round duration (see \cref{fig:roundvspfa,fig:roundvspd}). A duration of three OFDM symbols achieves acceptable performance ($P_{fa} \approx 10^{-4}$, $P_d \approx 0.999$), whereas shorter durations lead to significantly higher error rates.

Specifically, one-symbol rounds provide insufficient energy integration, resulting in poor detection performance ($P_d \approx 0.5$). Two-symbol rounds improve detection but exhibit elevated false alarm rates due to limited robustness against noise and spectral leakage. A duration of three OFDM symbols provides the best trade-off, ensuring reliable detection while allowing multiple contention rounds within a single slot. This configuration is adopted as the baseline for the hardware implementation of the \ac{RCS} algorithm.

\subsection{Experimental results}
\label{subsec:results_exp}

\begin{figure}[t]
\centering\includegraphics[width=0.32\textwidth]{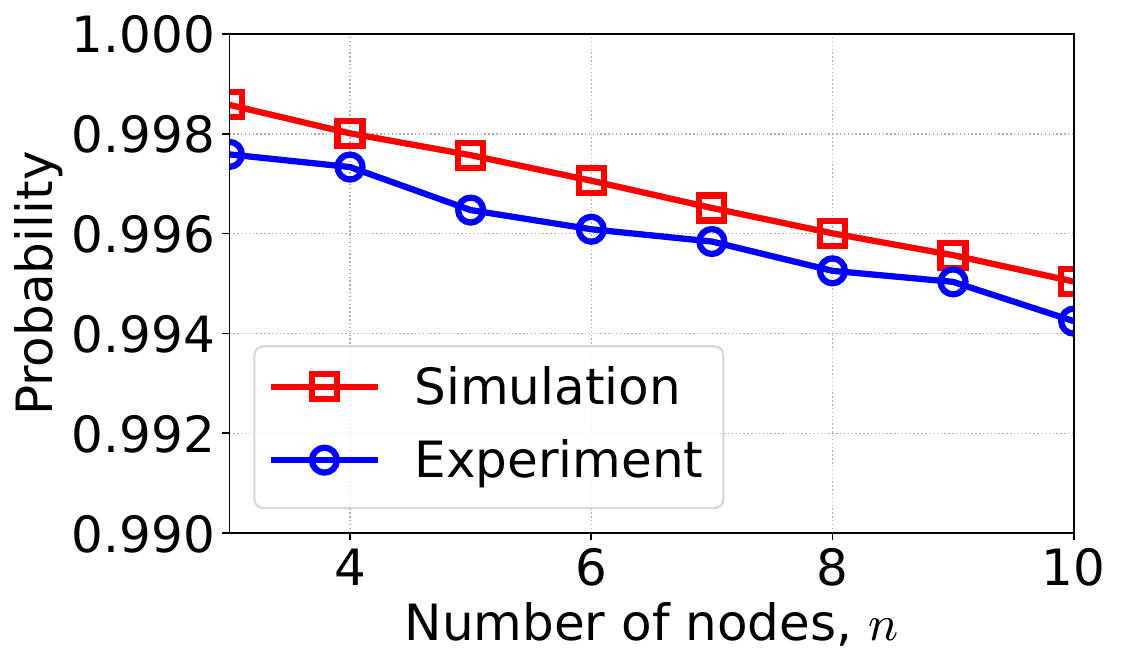}
\caption{Comparison between experimental and simulation-based success probability.}
\label{fig:exp_sim_psucc}
\vspace{-0.5cm}
\end{figure}

The experimental success probability, compared with simulation results, is shown in \cref{fig:exp_sim_psucc}. Please note that the simulation parameters shown in \cref{fig:exp_sim_psucc} are consistent with the experimental parameters reported in \cref{tab:experiment_parameters}.
The comparative analysis reveals that the experimental $P_{succ}$ measured on the USRP nodes is slightly lower than the values obtained through simulations. This performance gap is directly attributable to the emergence of $P_{nowin}$ within the hardware based setup. While the simulation environment assumes ideal conditions where every non-colliding contention results in a success, the physical implementation introduces real-world impairments. Specifically, a small fraction of these contention slots is lost due to false alarms during the sensing phase. In such instances, environmental impulse noise or interference is erroneously interpreted as an active tone from a competitor, leading nodes to defer and resulting in an inconclusive contention cycle even in the absence of actual collisions. 

\begin{figure}[t]
\centering\includegraphics[width=0.32\textwidth]{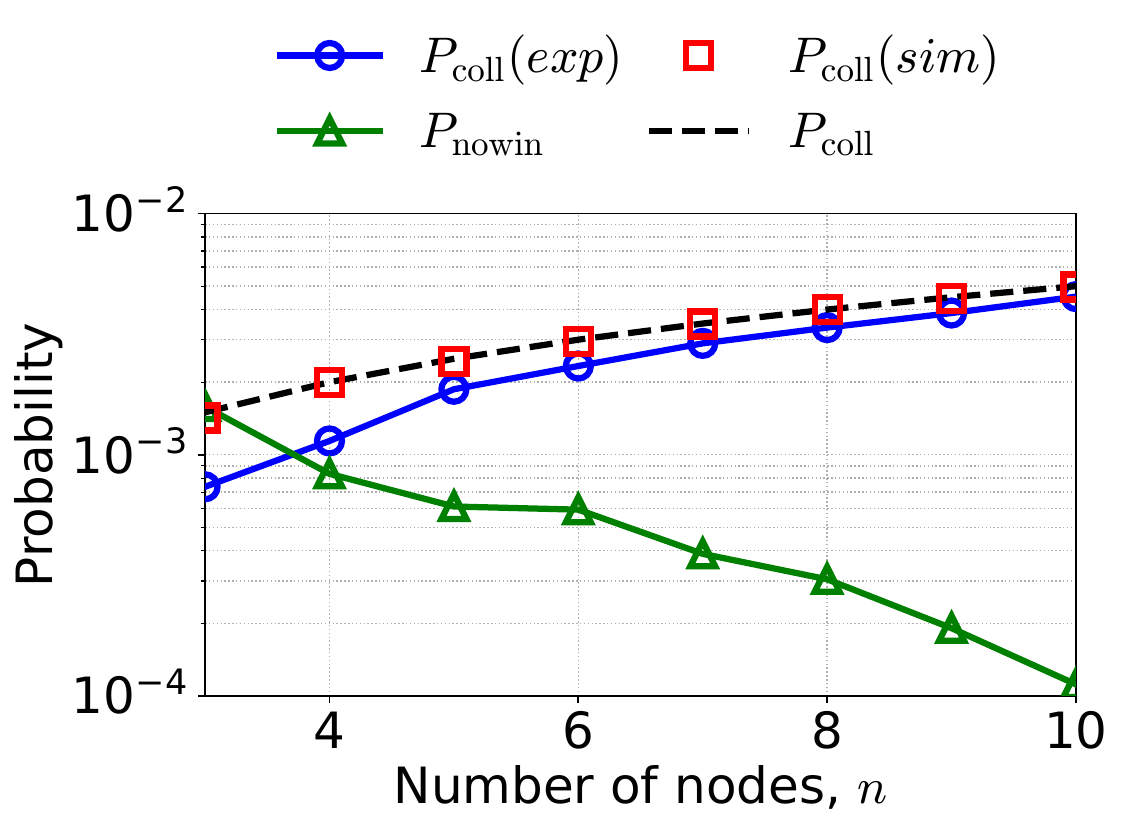}
\caption{Comparison of experimental and simulation-based collision probabilities, the theoretical upper bound on collision probability, and the no-winner probability.}
\label{fig:exp_coll_nowin}
\vspace{-0.5cm}
\end{figure}

The no-winner probability is illustrated in \cref{fig:exp_coll_nowin}, which compares simulation and experimental results. The figure reports the theoretical upper bound on collision probability from \cref{eq:pcupperbound2} (black curve), the simulated collision probability (red markers), and the experimentally measured collision (blue curve) and no-winner (green curve) probabilities.

First, the presence of no-winner events ($P_{\mathrm{nowin}}$) reduces the number of contention phases that can result in collisions, as some rounds terminate prematurely due to sensing errors. 
Second, physical-layer impairments may induce a collision-to-success transition. When two nodes select the same sub-carrier in the final round, a collision is expected; however, sensing errors or noise fluctuations may cause one node to detect a spurious lower-frequency tone and withdraw. As a result, the other node gains access to the resource, effectively converting a potential collision into a successful transmission. Although infrequent, this effect further reduces the observed $P_{\mathrm{coll}}$ compared to the ideal model.

The experimental results indicate that the no-winner contention probability ($P_{nowin}$) decreases as the number of contending nodes increases. This trend can be attributed to the fact that a higher density of participants increases the likelihood that false detections caused by noise are masked by the actual tones transmitted by the stations. In a scenario with a limited number of available sub-carriers, the event in which a station erroneously withdraws due to a low index false tone becomes increasingly rare as the spectrum becomes more populated. 
Conversely, the collision probability ($P_{coll}$) scales proportionally with the number of contending nodes, as the statistical likelihood of multiple stations selecting an identical sub-carrier in the final round increases. This inverse relationship between $P_{nowin}$ and $P_{coll}$ is depicted in \cref{fig:exp_coll_nowin}, which shows how the growth of $n$ shifts the system's primary impairment from sensing-induced deferrals to physical collisions. The experimental data reveal a clear transition in the dominant impairment factor as the network density increases.

\section{Conclusion}
\label{sec:conclusion}

This paper introduced \ac{RCS}, a novel resource allocation algorithm for NR-V2X and towards 6G sidelink communications that departs from \ac{DS} and {SPS}. By replacing reservation-based mechanisms with a multi-round contention process, \ac{RCS} enables fully distributed, feedback-driven resource selection, mitigating persistent collisions and adapting to dynamic vehicular environments. 
Simulation results demonstrate that \ac{RCS} consistently outperforms \ac{DS} and \ac{SPS} in terms of reliability and timeliness, achieving higher \ac{PRR}, lower loss rates, and improved \ac{PIR} and \ac{AoI}, particularly under high load conditions.
The feasibility of \ac{RCS} is further validated through an \ac{SDR}-based proof-of-concept. Despite hardware impairments, experimental results align with theoretical and simulation trends, confirming that \ac{RCS} operates reliably in practical settings and represents a viable solution for decentralized resource allocation in 5G NR and beyond.

Future work will focus on detailed modeling of the proposed \ac{RCS} algorithm under more realistic conditions, including aperiodic traffic, hidden nodes, signal propagation and fading models, as well as experimental investigations of scenarios involving multiple winners in the contention phase.

\section*{Acknowledgments}

This work was partially funded by Sapienza University of Rome under the “Progetti per Avvio alla Ricerca – Tipo 2” program (2025) for the project 5GSL-RTA-VRUP “Advancing 5G Vehicular Communication for Real-Time Awareness and VRU Protection” (prot. AR225199B968B6C6)

\printbibliography

\end{document}